# Next Generation of Ultra-Coarse-Graining: Self-Consistent Inference of Critical Internal States

Weizhi Xue, Xiao Shuai, Gregory A. Voth*

Department of Chemistry, Chicago Center for Theoretical Chemistry, James Franck Institute, and Institute for Biophysical Dynamics, The University of Chicago, Chicago, IL 60637, USA

*Corresponding Author: gavoth@uchicago.edu

## Abstract

Bottom-up coarse-graining can systematically expand the length and effective time scales accessible to molecular dynamics (MD) simulations, but the information loss during coarse-graining can prevent an accurate incorporation of the multistate phenomena common in complex biomolecular dynamics. The Ultra-Coarse-Graining (UCG) method projects the influence of discrete "internal states" onto the coarse-grained (CG) mapped molecules. These implicit, "quantum-like" extended degrees of freedom can significantly extend the expressiveness of the CG model. Previous UCG methods relied either on explicit Monte-Carlo-like sampling of such internal states or a mixed ensemble approach in the rapid local equilibrium (RLE) limit. In RLE-UCG, the internal state updates depend on user-defined collective variables (CVs, e.g., local density), and a local approximation that will break down for correlated internal states within and between CG molecules. We present here Self-Consistent UCG (SC-UCG), an integrated methodology that utilizes the underlying UCG interactions directly for internal state designation, without designing collective variables in the CG ensemble. For simulations, the internal state probabilities are self-consistently determined via message-passing in simulation time. We also enhance the RLE Hamiltonian with the Bethe approximation plus AI-based inference to allow explicit internal state correlations between UCG beads. For forcefield training, we develop Multilayer Internal State Consistency (MISC), a machine-learning-based method derived from the relative entropy minimization principle that does not require iterative sampling of the intermediate forcefields. We then apply SC-UCG to a model tetramer system with a second-order symmetry-breaking phase transition from the supercritical racemic fluid into subcritical D-rich and L-rich fluids. We show that SC-UCG can capture collective switching events of internal UCG states in the subcritical region, and it exhibits temperature transferability by recapitulating the phase transition, even if trained from a single-temperature dataset.

## I. Introduction

Molecular Dynamics (MD) is a well-established sampling technique for simulation of material, chemical, and biomolecular systems.[1] However, its computational capacity is constrained by the computational resources available. Coarse-grained (CG) forcefields (FFs) provide a systematic way to directly reduce the number of integrable degrees of freedom (DOFs).[2–7] Bottom-up CG methods, including the Multiscale Coarse Graining (MS-CG),[8–10] the Iterative Boltzmann Inversion (IBI),[11,12] and the Relative Entropy Minimization (REM) framework,[13–16] all aim to derive such CG FFs from a rigorous perspective, solely from the microscopic all-atom data. On the other hand, top-down CG methods, such as the Martini approach[6,7], typically fit CG FFs from experimental data or other observables. An inevitable feature of all CG methods is some loss of more detailed information, which can lead to the following problems: (1) CG FFs trained for one species may not apply to another species (molecular transferability); (2) CG FFs trained at one temperature and pressure condition may not apply to other temperatures or pressures (temperature or thermodynamic transferability); (3) a loss of some realistic properties and "missing entropy"[17–21] from the fast DOFs that have been effectively integrated out; and (4) for inherently multistate behavior in a particular system, the CG distribution may lose its ability to correctly sample some states or sample them not at all (tied to the "representability problem").[4,22]

The Ultra-Coarse-Graining (UCG) approach was introduced to address problem (4) and partly problem (3) by introducing auxiliary variables into the CG forcefield that represent select internal degrees of freedom.[23–26] Sometimes the CG atomistic system can contain certain key features – such as torsional angles not explicitly resolved in the coarse-grained molecule – then these may in turn give rise to slowly-transitioning DOFs. In these cases, the UCG approach preserves the effects of those DOFs in the form of discrete-valued auxiliary variables, also referred to as "internal states," having a certain isomorphism with quantum mechanics. UCG particles thus possess a discrete DOF under this setting, and a particle in internal state $s_i$ usually interacts with other particles differently when they are in another internal state $s_i' \neq s_i$. The model distribution for the CG system is no longer a single Boltzmann ensemble under a single CG forcefield, but rather a mixed ensemble from different effective interactions.[24,25] This approach is theoretically connected with mixed-ensemble free energy sampling techniques such as Integrated Hamiltonian Sampling[27] and Integrated Boltzmann Sampling,[28–30] and have been applied in free energy calculations.

To date, two limiting cases for UCG modeling have been developed in our group. For the slow or rare state transition limit,[24] Monte-Carlo switching of the discrete internal states has been applied to rare events, e.g., for large biomolecules, such as the ATP/ADP nucleotide state changes in actin filaments[31] and the inactive/active conformational selection of the HIV-1 virus capsid proteins during capsid assembly.[32,33]

The slow transition UCG approach has also been applied to simpler interacting CG liquid systems such as 1,2-dichloromethane to illustrate the approach.[24] In such cases, the transitions between internal UCG states happen as rare events, which correspond to anti-gauche torsional transitions at the all-atom level. The effective UCG state switches happen on a much larger timescale than a CG MD timestep, during which the faster orthogonal DOFs can relax.[23,24] We shall refer to this method as Monte-Carlo UCG (MC-UCG).

The UCG internal state fast-transition limit, on the other hand, treats the internal state of a CG site as a fast DOF that exists in local quasi-equilibrium with its neighboring sites.[25,26] In these scenarios, instead of having discrete internal state labels, each UCG particle exists in a mixed state according to some internal state probability (ISP) function. The orthogonal DOFs evolve under a free energy surface of these internal states, conditioned on other DOFs. This method is called UCG with rapid local equilibrium (RLE-UCG). RLE-UCG has been successfully applied in modeling liquid-gas equilibrium and self-assembly of the lipid bilayer of 1,2-dipalmitoyl-sn-glycero-3-phosphocholine (DPPC). The REM approach[13–16] has also been extended to accommodate UCG modeling, especially for RLE-UCG.[26]

As powerful as MC-UCG and RLE-UCG have proven to be so far, they have only been applied to systems in which the internal states have little or no correlation with other internal states in other UCG sites. Correlation-induced cooperativity exists, among other possibilities, for conformational changes, cooperative chemical reactions, and phase transitions, such as the ripple phase behavior in biological membrane systems[34–37] and chiral symmetry breaking in certain organic materials.[38,39] For example, a tetramer model system has been demonstrated to exhibit a chiral-racemic transition in its fluid phase, in which the transition of chiral states of each molecule is highly correlated across the system in the subcritical region.[40–42] This critical transition falls within the 3D Ising universality class. We will show here that such phase transitions can be described by the MC-UCG approach if adequately sampled, as it directly samples the internal state as a discrete DOF. The current RLE-UCG approach, however, relies on two premises that prevents it from sampling such phenomena. First, the ISP functions are functions of CG collective variables (CVs) – which means the current RLE-UCG forcefield is geometry-based – and contains its own parameters to optimize. This assumes that the identification of the internal state of a CG site in the atomistic ensemble is inherently dependent on neighboring CG sites, so the internal states are no longer "internal" – they are "external". The Internal State Regression (ISR) method[26] proposes a systematic way for the optimization of extra parameters in the ISP functions, but it still relies on geometry-based CVs. A more serious problem of such CV selection is that it may break the symmetry of the CG forcefield if the broken symmetry in the configurations arises from an all-atom potential with such symmetry. Second, RLE-UCG introduces a very simple mean-field (MF) approximation to the effective Hamiltonian by decoupling the joint distribution of all internal states in the system into products of individual ISP functions. This eliminates the probabilistic correlation between pairs or cliques of UCG particles. Therefore, although RLE-UCG is good at describing well-defined phase separation systems exhibiting low fluctuations,[43] such as the liquid-gas interface, it is currently unable to handle phase transitions and more highly correlated collective fluctuations in internal UCG state space without special model tuning.

In this work, we introduce the Self-Consistent UCG (SC-UCG) method, which aims to address these two limitations in the current RLE-UCG approach and provide a unified framework for UCG modeling in both slow-mixing and fast-mixing limits. In line with MC-UCG, we treat the internal states as discrete-valued degrees of freedom that do not explicitly depend on hand-selected geometric collective variables or descriptors from its neighboring environment. In line with RLE-UCG, we adopt the mixed Hamiltonian formalism and derive, via variational principle, the definition of ISP of a CG site in the CG ensemble, which naturally results in a message-passing form in Graph Neural Networks (GNNs),[44–46] with the UCG forcefield directly acting as trainable weights and the inverse temperature $\beta$ controlling the steepness in internal state assignment. In other words, we let the UCG interactions directly determine the internal state probabilities in the UCG ensemble, so a choice of a CV is no longer required in the CG ensemble. We further enhance the representability of the RLE-UCG Hamiltonian by introducing explicit pairwise internal state correlations with the Bethe-Peierls approximation.[47,48] For forcefield training, we derived the Multilayer Internal State Consistency (MISC) training framework directly from the relative entropy framework on the joint distribution of positions and internal states. MISC is a self-supervised learning method that optimizes forcefields through graph backpropagation on the atomistic data, and it does not require iterative sampling of CG forcefields in each iteration of optimization.

We then apply the SC-UCG simulations and the MISC training framework to a chiral-racemic phase transition system.[41] We show that SC-UCG simulation is capable of capturing large-scale, system-wide fluctuations and correlations in internal states near the critical point. We show that, even if the interactions are learned from one temperature only, the SC-UCG model can describe phase transitions in internal state space due to its natural inclusion of temperature and forcefields in the ISP function, and it exhibits temperature transferability even if the forcefield training is only conducted for one temperature.

## II. Theory and Methodology

### A. Self-Consistent Inference of Internal States in Ultra-Coarse-Graining

We assume the all-atom (AA) configurations are sampled from the equilibrium Boltzmann distribution $p_{AA}(r^n)$, derived from the all-atom potential $u_{AA}(\boldsymbol{r}^n)$, such that $p_{AA}(\boldsymbol{r}^n) \propto e^{-\beta u_{AA}(\boldsymbol{r}^n)}$, where $\beta = \frac{1}{k_B T}$.

In bottom-up coarse-graining (CG), we use a mapping operator $M$ that maps all-atom coordinates $\boldsymbol{r}^n$ to CG coordinates $\boldsymbol{R}^N$, and aim to derive, from the obtained AA data, a CG forcefield, $u_{CG}(\boldsymbol{R}^N)$, that acts upon this reduced set of CG coordinates. More generally, we aim to obtain a CG model distribution $p_{CG}(\boldsymbol{R}^N)$, as close to the all-atom distribution, projected onto the CG maps, as possible. Therefore, the most generalized bottom-up CG objective writes as

$$\boldsymbol{\theta}^* = \text{argmin}_{\boldsymbol{\theta}}\, D[p_{AA}(\boldsymbol{r}^n)||p_{CG}(M(\boldsymbol{r}^n)|\boldsymbol{\theta})] \tag{1}$$

where $D$ represents a generalized distance metric. It can be the Kullback-Leibler (KL) divergence used in REM, or the Fisher divergence used in Force-Matching. Usually, when one single CG forcefield $u_{CG}(\boldsymbol{R}^N)$ is concerned, we also assume Boltzmann sampling with this CG forcefield: $p_{CG}(\boldsymbol{R}^N) \propto e^{-\beta u_{CG}(\boldsymbol{R}^N)}$, so that normal MD simulation with the CG forcefield should recover the all-atom distribution projected onto the CG variables.

When the all-atom data is inherently multimodal, the representability given by one single CG forcefield, if that forcefield is approximate, may not suffice. The UCG theory broadly assumes that the CG distribution is a *mixed ensemble* (or *integrated ensemble*, or *extended ensemble*) along some (possibly abstract) additional degrees of freedom with an index, $\sigma$, such that

$$p_{CG}(\boldsymbol{R}^N) \propto \sum_{\sigma} e^{-\beta u_{CG}(\boldsymbol{R}^N,\sigma)} \tag{2}$$

These additional degrees of freedom are called "internal states" in the UCG approach. They can either be possessed by each CG particle or by groups of CG particles (such as whole molecules). The effective CG interaction in the RLE limit of UCG theory is now a conditional free energy:

$$u_{\text{mix}}(\boldsymbol{R}^N) = -\frac{1}{\beta}\log\sum_{\sigma} e^{-\beta u_{CG}(\boldsymbol{R}^N,\sigma)} \tag{3}$$

In a mixed ensemble, one can define the posterior distribution of these additional degrees of freedom:

$$p(\sigma|\boldsymbol{R}^N) = \frac{e^{-\beta u_{CG}(\boldsymbol{R}^N,\sigma)}}{\sum_{\sigma} e^{-\beta u_{CG}(\boldsymbol{R}^N,\sigma)}} \tag{4}$$

Using thermodynamic properties, the conditional free energy $u_{\text{mix}}(\boldsymbol{R}^N)$ can be decomposed into its (conditional) enthalpic and (also conditional) entropic contributions:

$$u_{\text{mix}}(\boldsymbol{R}^N) = \langle E(\boldsymbol{R}^N)\rangle - TS(\boldsymbol{R}^N) = \sum_{\sigma} p(\sigma|\boldsymbol{R}^N)u_{CG}(\boldsymbol{R}^N,\sigma) + k_B T\sum_{\sigma} p(\sigma|\boldsymbol{R}^N)\log p(\sigma|\boldsymbol{R}^N) \tag{5}$$

It can be shown that the definition of the posterior in Eq. (4) is a variational optimum for Eq. (5) if we variationally minimize $u_{\text{mix}}(\boldsymbol{R}^N)$ with the constraint $\sum_{\sigma} p(\sigma|\boldsymbol{R}^N) = 1$.

For per-particle scenarios in which each UCG particle $i$ possesses its own internal state over $q$ possible states in total, as the system size increases, the number of configurations of $\sigma = (s_1, \dots, s_N)$ becomes exponentially large, often prohibiting efficient sampling and leading to extreme computational demands if one were to naively try to explicitly enumerate all of the UCG states. However, in such cases, simplification to the posterior, $p(\sigma|\boldsymbol{R}^N)$, is needed. In UCG-RLE, it is often assumed that the internal state of each particle is only immediately affected by its neighbors, and they adjust rapidly in response to fluctuations in neighboring environments to maintain local equilibrium. Therefore, a mean-field type decomposition can be used:

$$p(\sigma|\boldsymbol{R}^N) = \prod_i p(s_i|\boldsymbol{R}^N) \tag{6}$$

Furthermore, assuming pairwise decomposability for the state-specific forcefield, given by

$$u_{CG}(\boldsymbol{R}^N, \sigma) = \sum_i h_{i,s_i}(\boldsymbol{R}^N) + \sum_{\langle i,j\rangle} u_{ij,s_i s_j}(R_{ij}) \tag{7}$$

we can write the mixed Hamiltonian in UCG-RLE as

$$u_{\text{mix}}(\boldsymbol{R}^N) = \sum_{i,s_i} p_i(s_i|\boldsymbol{R}^N)\left[h_{i,s_i} + \frac{1}{\beta}\log p_i(s_i|\boldsymbol{R}^N)\right] + \sum_{\langle i,j\rangle}\sum_{s_i s_j} p_i(s_i|\boldsymbol{R}^N)p_j(s_j|\boldsymbol{R}^N)\, u_{ij,s_i s_j}(R_{ij}) \tag{8}$$

This expression is a mean-field free energy functional integrated over internal states, conditioned on the system configuration. For more compact notation, for each particle $i$, we represent the single-particle **internal state probability (ISP)** functions with a vector $\boldsymbol{p}_i$; likewise, we have the chemical potentials with vector $\boldsymbol{h}_i$ and UCG pairwise interaction matrix $\boldsymbol{U}_{ij}$. We can therefore write this mixed Hamiltonian in a matrix form

$$U_{\text{RLE}} = \sum_i \boldsymbol{p}_i^{\text{T}}(\boldsymbol{h}_i + \frac{1}{\beta}\log \boldsymbol{p}_i) + \sum_{\langle i,j\rangle} \boldsymbol{p}_i^{\text{T}}\boldsymbol{U}_{ij}\boldsymbol{p}_j \tag{9}$$

Up to this point, no new methods have been established. It is useful to point out that in the current UCG-RLE implementations, the ISP functions are designed functions dependent on real space CVs in both AA space and CG space (see Fig. 1a). For any phenomenon of interest, such as liquid-gas phase separation, one needs to first identify the CV in CG space, select a form for the ISP functions, and train, preferably via Internal State Regression (ISR), the additionally introduced parameters in these ISP functions. Such CV dependence may break the desired symmetry implied in the CG model if the selected CV contains noninvariant components to symmetry operations. For example, defining the CG CV in terms of its angle to the $z$-axis direction introduces $z$-axis dependence into the ISP and thus the CG forcefield, which breaks the rotational symmetry of the forcefield (although the induced physical phenomena can and often have broken symmetry). Moreover, because of the CV dependence, the force in such simulations contains additional ISP gradients $\nabla_{\boldsymbol{R}}\boldsymbol{p}_i(\boldsymbol{R})$, which adds an additional computational cost. Therefore, removing such CV dependence from the CG model is desired.

Hence, following the Lagrange variational method in Equations (4) and (5), we employ variational minimization of $U_{\text{RLE}}$ with respect to the ISP vector $\boldsymbol{p}_i$ with a normalization constraint can be performed with an Lagrangian multiplier $L$, given by

$$L = U_{\text{RLE}} - \sum_i \lambda_i\left(\sum_{s_i} p_{i,s_i} - 1\right) \tag{10}$$

$$\frac{\delta L}{\delta p_{i,s_i}} = \frac{\delta L}{\delta \lambda_i} = 0, \forall i \tag{11}$$

This yields a self-consistent equation (SCE) in a message-passing (MP) form:

$$\boldsymbol{p}_i = \sigma_\beta\left(\boldsymbol{h}_i + \sum_{j\in N(i)} \boldsymbol{U}_{ij}\boldsymbol{p}_j\right) \tag{12}$$

where we define $[\sigma_\beta(x)]_i = \frac{e^{-\beta x_i}}{\sum_j e^{-\beta x_j}}$ as the softmax function for an arbitrary vector $\boldsymbol{x}$, with an inverse temperature $\beta = 1/k_B T$, in correspondence with the regular vector-valued softmax function $[\sigma(x)]_i = \frac{e^{-x_i}}{\sum_j e^{-x_j}}$. In two-state scenarios and assuming $p_i = p_j = \bar{p}$, one immediately recovers the classical mean-field solution of the Ising model. Of course, in the UCG scenario, we do not need any further simplifications to this form to prevent limiting its applicability.

It is particularly useful to understand this framework in the terms of Graph Neural Networks (GNNs).[49,50] In fact, each iteration of the self-consistent equation:

$$\boldsymbol{p}_i^{(t+1)} \leftarrow \sigma_\beta \left( \boldsymbol{h}_i^{(t)} + \sum_{j \in N(i)} \boldsymbol{U}_{ij}^{(t+1)} \left( R_{ij} | \boldsymbol{\theta} \right) \boldsymbol{p}_j^{(t)} \right) \tag{13}$$

represents a message-passing GNN (MP-GNN) layer. The *nodes* are particles, the *edge* connections are given by the neighbor list, and the *trainable weights* are directly the forcefield parameters. The total energy of the system is given by weighted summation over node and edge features. The *message* from particle $j$ to $i$ is the averaged interaction $\boldsymbol{U}_{ij}\boldsymbol{p}_j$ it exerts onto each possible internal state of $i$; the *aggregation* method is the summation; the *node update* function is a temperature-dependent sigmoid activation. We shall denote one such layer as one **Internal State Convolution (ISC)** layer, and we rewrite the update from $n$ iterations as $\left\{p_i^{(n)}\right\} \leftarrow \sigma^n \left( \left\{p_i^{(0)}\right\} \middle| \boldsymbol{\theta} \right)$, where we denote all parameters to learn as $\boldsymbol{\theta}$. In ISC layers, each trained UCG forcefield directly acts as scores for node classification. Therefore, we can also call this architecture an **Internal State Classifier (ISC)**. A generalized graphical structure is depicted in Fig. 1a.

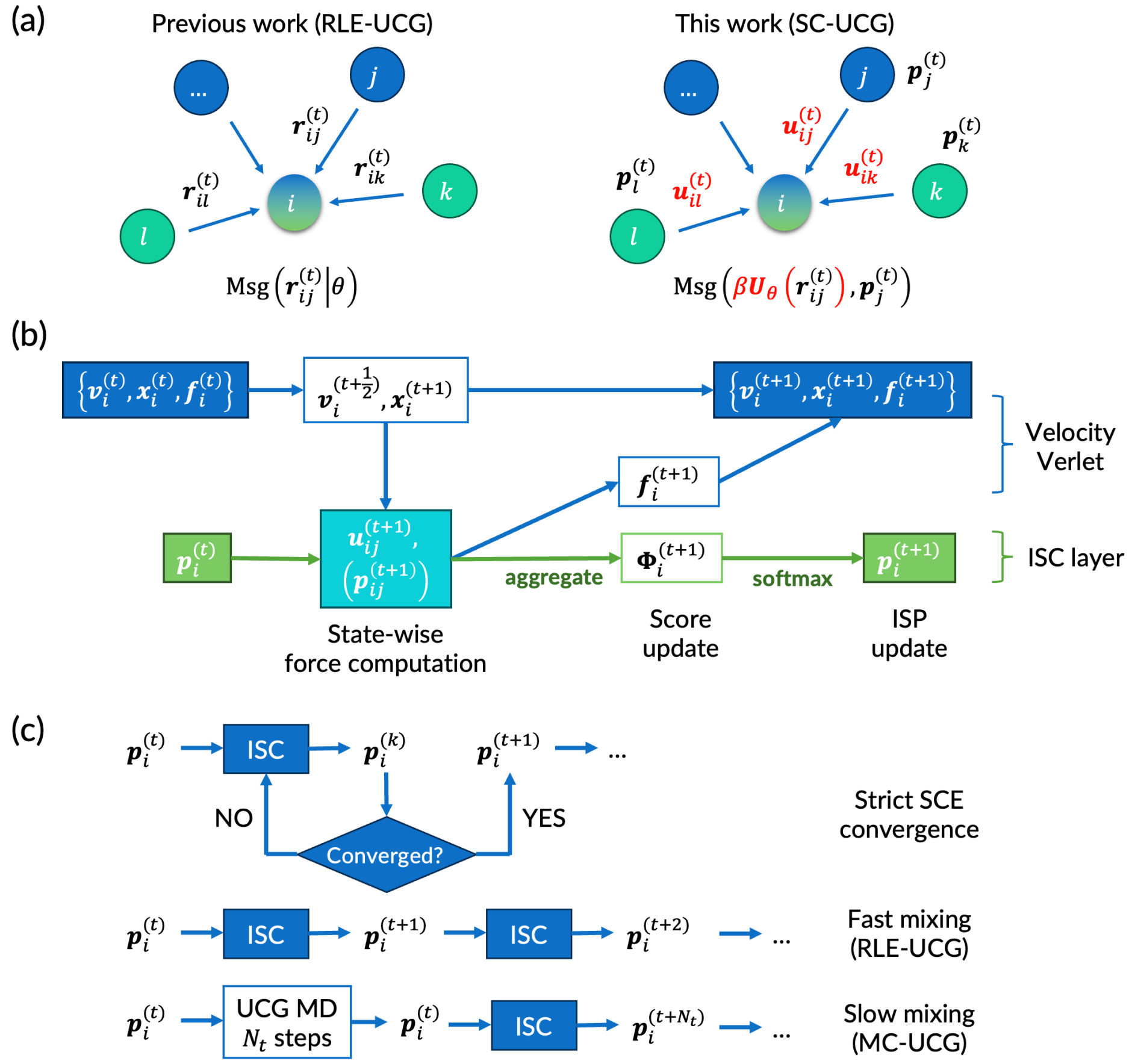

**FIG. 1. Overview of SC-UCG simulation.** (a) Message construction in SC-UCG compared with RLE-UCG. RLE-UCG directly uses purely geometric edge attributes $\{r_{ij}\}$ for message passing, while SC-UCG uses the state-wise interaction matrix $\left\{\boldsymbol{u}_{ij}\left(r_{ij}^{(t)}\right)\right\}$ and node probabilities of neighbors $\left\{\boldsymbol{p}_j^{(t)}\right\}$ for message passing. (b) What a timestep of simulation looks like in SC-UCG simulations. Inference of internal states by the ISC layer works in tandem with the integration scheme (e.g. Velocity Verlet here). (c) SC-UCG unifies different internal state update regimes in current UCG methods. For strict SCE convergence, each internal state probability (ISP) update requires multiple ISC passes until SCE convergence, before the simulation can move on to the next timestep. For fast mixing dynamics such as in RLE-UCG, 1 layer of ISC suffices within one MD timestep. For slow mixing dynamics where internal state transitions in individual UCG particles are rare events, one can allow $N_t$ MD relaxation before next ISC pass.

During UCG simulation, for perfect self-consistent convergence, one would expect multiple iterations within one MD step for strict self-consistent convergence. Doing so, however, adds an extra burden on processor communication and computation. We instead stack the ISC layers in the arrow of time directly during simulations (see Fig. 1b). In practice, message aggregation is conducted during pairwise force computation, and the node update is performed after force computation.

Stacking too many ISC layers for SCE convergence is known to produce "over-smoothing problem" in deep MP-GNNs[49]. Depending on the specific system of interest, the "over-smoothing" problem may not be a problem in realistic systems – but rather it embodies the exact "rapid local equilibrium" the RLE-UCG method itself assumes and intends to simulate. Usually, the SCE convergence is typically much faster than significant real space configuration changes in MD, so our use of 1 ISC layer per MD step falls within the fast-mixing RLE category. On the other hand, one can also simulate the slow mixing limit[24] for the internal state updates within the SC-UCG framework, by using one ISC pass after a *tunable* number $n_t$ of timesteps (see **Fig. 1c**). If one uses the Metropolis algorithm to sample back the discrete space, one returns to the Monte-Carlo based UCG method.

This variational framework also allows for constrained inference. It is also convenient to enforce constraints with additional Lagrange multipliers. One common constraint is the material or charge balance, when, for example, simulating proton transport in water and the total number of hydrated protons should be kept fixed. In this case, the Lagrangian objective is

$$L = U_{RLE} - \sum_i \lambda_i \left( \sum_{s_i} p_{i,s_i} - 1 \right) - \sum_s \mu_s \left( \sum_i p_{i,s_i} - \nu_s \right) \tag{14}$$

The SCE iteration will be a set of coupled equation for $\{\boldsymbol{p}_i\}$ and $\boldsymbol{\mu}$:

$$\begin{cases} \boldsymbol{p}_i = \sigma_\beta(\boldsymbol{\Phi}_i - \boldsymbol{\mu}) \\ \sum_i \sigma_\beta(\boldsymbol{\Phi}_i - \boldsymbol{\mu}) = \boldsymbol{\nu} \end{cases} \tag{15}$$

where $\boldsymbol{\Phi}_i^{(t)} = \boldsymbol{h}_i^{(t)} + \sum_{j \in N(i)} \boldsymbol{U}_{ij}\left(\boldsymbol{R}^N(t)\right) \boldsymbol{p}_j^{(t)}$. We can see that the Lagrange multiplier $\boldsymbol{\mu}$ acts as a varying global chemical potential for the internal states, on top of the individual chemical potentials $\{\boldsymbol{h}_i\}$. This gives rise to a three-step SCE iteration scheme: (1) Obtain $\boldsymbol{\Phi}_i^{(t)}$ during pairwise force calculation; (2) Update $\boldsymbol{p}_i^{(t+1)} = \sigma_\beta\left(\boldsymbol{\Phi}_i^{(t)} - \boldsymbol{\mu}^{(t)}\right)$ after force calculation; (3) solve or determine $\boldsymbol{\mu}^{(t+1)}$ such that $\sum_i \sigma_\beta\left(\boldsymbol{\Phi}_i^{(t)} - \boldsymbol{\mu}^{(t+1)}\right) = \boldsymbol{\nu}$ after force calculation.

## B. Extending RLE-UCG with Bethe Free Energy Functionals

Recognizing the mean-field nature of the RLE approximation, we can enhance the UCG-RLE formalism directly with other free energy approximations and also employ variational inference of internal states for them. In this paper, we use the Bethe-Peierls approximation[47,48,51] as an example, to introduce explicit correlation terms in the free energy functional while keeping its pairwise separability. While we note that there exists more accurate free energy approximation and inference methods containing three-body or more probability correlations, such as the Cluster Variation Method (CVM),[52] Adaptive Cluster Expansion (ACE),[53–55] etc., these multi-body correlations come at the cost of simulation speed. We therefore prefer to preserve a computationally inexpensive pairwise summation form. Hence, instead of approximating $\boldsymbol{p}_{ij} = \boldsymbol{p}_i \boldsymbol{p}_j^{\mathrm{T}}$ for $\forall \langle i,j \rangle$, we explicitly consider the pairwise contribution to the conditional entropy. We thus change the entropic contribution to $u_{mix}$ in (8) into the Bethe entropy:

$$S_{Bethe} = \sum_i (1 - d_i) \boldsymbol{p}_i^T \log \boldsymbol{p}_i + \sum_{\langle i,j \rangle} \boldsymbol{p}_{ij} * \log \boldsymbol{p}_{ij} \tag{16}$$

where we use $A * B = \sum_{ij} A_{ij} B_{ij}$ to represent the Frobenius product of matrices. The resulting CG effective interaction is written as

$$U_{\text{Bethe-RLE}} = \sum_i \boldsymbol{p}_i^{\mathrm{T}} \left[ \boldsymbol{h}_i + \frac{1}{\beta}(1-d_i)\log \boldsymbol{p}_i \right] + \sum_{\langle i,j \rangle} \boldsymbol{p}_{ij} * \left( \boldsymbol{U}_{ij} + \frac{1}{\beta}\log \boldsymbol{p}_{ij} \right) \tag{17}$$

For the generalized $q$-state scenario, the well-known Belief Propagation[48,56] algorithm is used for inference, and can be derived from variational minimization of $U_{\text{Bethe}}$ under these normalization requirements for the matrices $p_{ij}$ to yield

$$\begin{gathered} p_i(s_i) = \sum_{s_j} p_{ij}(s_i, s_j)\,;\; p_j(s_j) = \sum_{s_i} p_{ij}(s_i, s_j) \\ \sum_i p_i(s_i) = 1;\, \sum_{s_i s_j} p_{ij}(s_i, s_j) \;= 1 \end{gathered} \tag{18}$$

For two-state UCG scenarios, these normalization requirements can be implicitly satisfied by treating $p_{ij}(s_i, s_j)$ as functionals of $p_i(s_i)$ and $p_j(s_j)$ and solving for its own variational statement. Let $p_i(1) = q_i$, $p_j(1) = q_j$, $p_{ij}(1,1) = \xi_{ij}$, and other $p_{ij}(s_i, s_j)$ can be derived from them:

$$\boldsymbol{p}_{ij} = \begin{pmatrix} p_{ij,00} & p_{ij,10} \\ p_{ij,01} & p_{ij,11} \end{pmatrix} = \begin{pmatrix} 1 + \xi_{ij} - q_i - q_j & q_j - \xi_{ij} \\ q_i - \xi_{ij} & \xi_{ij} \end{pmatrix} \tag{19}$$

Then, the variational statement can be expressed as

$$\frac{\partial U_{\text{Bethe-UCG}}}{\partial \xi_{ij}} = \beta J_{ij} - \ln \frac{\xi_{ij}(1 + \xi_{ij} - q_i - q_j)}{(q_i - \xi_{ij})(q_j - \xi_{ij})} = 0 \tag{20}$$

where $J_{ij} = U_{ij}(0,0) + U_{ij}(1,1) - U_{ij}(0,1) - U_{ij}(0,0)$ is the effective coupling of the system. This gives a closed functional form for $\xi_{ij}$, dependent on functions $q_i, q_j, J_{ij}$:

$$\xi_{ij}^*(q_i, q_j, J_{ij}) = \frac{Q_{ij} - \sqrt{Q_{ij}^2 - 4\alpha_{ij}(1 + \alpha_{ij}) q_i q_j}}{2\alpha_{ij}} \tag{21}$$

where $\alpha_{ij} = e^{-\beta J_{ij}} - 1$, $Q_{ij} = 1 + \alpha_{ij}(q_i + q_j)$. This results in the following edge update format:

$$\begin{gathered} \xi_{ij}^{(t+1)} \leftarrow \xi_{ij}^* \left( q_i^{(t)}, q_j^{(t)}, J_{ij}^{(t+1)} \right) \\ p_{ij}^{(t+1)} \leftarrow \begin{pmatrix} 1 + \xi_{ij}^{(t+1)} - q_i^{(t)} - q_j^{(t)} & q_j^{(t)} - \xi_{ij}^{(t+1)} \\ q_i^{(t)} - \xi_{ij}^{(t+1)} & \xi_{ij}^{(t+1)} \end{pmatrix} \end{gathered} \tag{22}$$

With the edge probabilities inserted, we can also perform variational optimization of $q_i$ and arrive also at a message-passing form of node update:

$$p_{i,s_i}^{(t+1)} = \sigma_{\beta,s_i} \left( h_{i,s_i} + \sum_{j \in N(i)} \sum_{sj} u_{ij,s_i s_j}^{(t+1)} \frac{p_{ij,s_i s_j}^{(t+1)}}{p_{i,s_i}^{(t)}} \right) \tag{23}$$

Therefore, the algorithmic structure is the same with MF except an additional edge update step is needed, in which one should sequentially compute $J_{ij}, \alpha_{ij}, Q_{ij}$, $\xi_{ij}^*$, and update $\boldsymbol{p}_{ij}$ *during* pairwise force computation before tallying and updating node ISP. In MP-GNNs, this Bethe-UCG approach amounts to assigning and updating edge embeddings and designing new message forms.

The above approach provides a generalizable framework for further exploration of probabilistic graph inference layers that uses UCG interaction matrices directly as edge features for internal state inference. For example, one could introduce the **attention** mechanism[57,58] by writing, from

$$\boldsymbol{p}_i = \sigma_\beta \left( h_{i,s_i} + \sum_{j \in N(i)} \alpha_{ij} \boldsymbol{u}_{ij} \boldsymbol{p}_j \right) \tag{24}$$

Or, from the Bethe approximation,

$$p_{i,s_i} = \sigma_{\beta,s_i} \left( h_{i,s_i} + \sum_{j \in N(i)} \sum_{sj} \alpha_{ij} u_{ij,s_i s_j} \frac{p^*_{ij,s_i s_j}}{p_{i,s_i}} \right) \tag{25}$$

where the attention score of neighbor $j$ to node $i$, $\alpha_{ij}$, is calculated via some softmax function over learnable edge score function $e_\phi$ with all neighbors:

$$\alpha_{ij} = \frac{e_\phi(\beta, u_{ij}, p_{ij}, R_{ij} \ldots)}{\sum_{k \in N(i)} e_\phi(\beta, u_{ik}, p_{ik}, R_{ik} \ldots)} \tag{26}$$

These edge attributes can include edge distance, edge ISP, etc. One can also make use of the multi-head attention mechanism to combine different types of correlation in the dataset. This formalism further appears in a fractional or edge-weighted Bethe approximations. Note that we can also rewrite the original Bethe update in a similar fashion, but with $\alpha_{ij,s_i s_j} = \frac{p_{ij,s_i s_j}}{p_{i,s_i} p_{j,s_j}}$ denoting a "cooperativity score" instead of "attention over neighbors".

### C. The Internal State Consistency Loss from the Relative Entropy Principle

Following previous discussions, it is also feasible to approach the inverse problem, i.e., learning the chemical potentials $\{\boldsymbol{h}_i\}$ and $\{\boldsymbol{U}_{ij}\}$, by developing the corresponding relative entropy minimization (REM) statement. The UCG model introduces the internal states as additional degrees of freedom in the model. These internal states can be observables, i.e., CVs of all-atom DOFs, in the all-atom dataset. This CV function can be scaled to the range of (0, 1) and thus acting as ground truth ISP labels $\boldsymbol{p}_i^{(0)}$ of mapped CG particles. The ISP labels can also be discrete (one-hot encoded internal state labels), representing the discretized internal state assignments in the model.

We begin by revisiting the relative entropy principle in UCG (UCG-REM)[26], which minimizes the relative entropy (KL divergence) between the CG joint distribution in position and internal state space, $p_{CG}(M(\boldsymbol{r}^n), \boldsymbol{s}^N | \boldsymbol{\theta})$, dependent on the CG forcefield parameters $\boldsymbol{\theta}$, and the joint distribution in AA dataset, $p_{AA}(\boldsymbol{r}^n, \boldsymbol{s}^N)$:

$$S_{\text{rel}} = \int \mathrm{d}\boldsymbol{r}^n \int \mathrm{d}\boldsymbol{s}^N \, p_{AA}(\boldsymbol{r}^n, \boldsymbol{s}^N) \log \frac{p_{AA}(\boldsymbol{r}^n, \boldsymbol{s}^N)}{p_{CG}(\boldsymbol{M}(\boldsymbol{r}), \boldsymbol{s}^N)} + S_{\text{map}} \tag{27}$$

where the joint distribution in the all-atom data space is defined via

$$p_{AA}(\boldsymbol{r}^n, \boldsymbol{s}^N) = p_{AA}(\boldsymbol{r}^n) p_{AA}(\boldsymbol{s}^N | M(\boldsymbol{r}^n)) \tag{28}$$

We can decompose the joint relative entropy in two terms: the conventional relative entropy of the marginal distribution over real space, and a conditional term:

$$S_{\mathrm{rel}} = \int \mathrm{d}\boldsymbol{r}^n p_{AA}(\boldsymbol{r}^n) \log \frac{p_{AA}(\boldsymbol{r}^n)}{p_{CG}(M(\boldsymbol{r}^n))} + \mathbb{E}_{\boldsymbol{r}^n \sim p(\boldsymbol{r}^n)}[S_{\mathrm{cond}}(\boldsymbol{r}^n)] \tag{29}$$

This condition term is written as

$$S_{\mathrm{cond}}(\boldsymbol{r}^n) = \int \mathrm{d}\boldsymbol{s}^N \, p_{AA}(\boldsymbol{s}^N | M(\boldsymbol{r}^n)) \log \frac{p_{AA}(\boldsymbol{s}^N | M(\boldsymbol{r}^n))}{p_{CG}(\boldsymbol{s}^N | M(\boldsymbol{r}^n))} \tag{30}$$

Since the UCG forcefield itself acts as a classifier to determine UCG internal state probabilities during simulations, and there is no CV dependence in SC-UCG that is used across AA and CG ensembles, we can no longer neglect the contribution of this term. Because each mapped UCG particle in the all-atom dataset has its own internal state, defined usually with its own internal degrees of freedom, and up to human prior knowledge of the all-atom system of interest, we write the expression

$$p_{AA}(\boldsymbol{s}^N | M(\boldsymbol{r}^n)) = \prod_i p_i^{(0)}(s_i | M(\boldsymbol{r}^n)) \tag{31}$$

In the similar spirit, because the dataset contains only node ISP, we also replace the CG internal state posterior $p_{CG}(\boldsymbol{s}^N | M(\boldsymbol{r}^n))$ with only node-level predictions from $K$ passes of ISC layers:

$$p_{CG}(\boldsymbol{s}^N | M(\boldsymbol{r}^n)) \rightarrow \prod_i p_i^{(K)}(s_i | M(\boldsymbol{r}^n), \boldsymbol{\theta}) \tag{32}$$

Therefore, we can simplify $S_{\mathrm{cond}}(\boldsymbol{r}^n)$ into the sum of node-level KL divergences such that

$$S_{\mathrm{cond}}^{(K)}(\boldsymbol{r}^n, \boldsymbol{\theta}) = \sum_i \sum_{s_i} p_i^{(0)}(s_i | M(\boldsymbol{r}^n)) \log \frac{p_i^{(0)}(s_i | M(\boldsymbol{r}^n))}{p_i^{(K)}(s_i | M(\boldsymbol{r}^n), \boldsymbol{\theta})} \tag{33}$$

The ISC layers require initial ISP input in order to make predictions. To ensure self-consistency in the model, this input should be the all-atom labels. The objective of forcefield training is to make the forcefield as close to self-consistency as possible, i.e., making the output as close to the input as possible. Therefore, we call the expectation of the conditional term as the **Internal State Consistency (ISC) loss**. In practice, only calculating the cross-entropy loss suffices for optimization, so we write the ISC loss as

$$S_{\mathrm{ISC}}^{(K)}(\boldsymbol{\theta}) = \mathbb{E}_{\boldsymbol{r}^n \sim p(\boldsymbol{r}^n)} \left[ -\sum_i \boldsymbol{p}_i^{(0)}(\boldsymbol{r}^n) \cdot \log \boldsymbol{p}_i^{(K)}(\boldsymbol{\theta} | M(\boldsymbol{r}^n)) \right] \tag{34}$$

Note that when $K = 1$ and $\boldsymbol{p}_i^{(0)}$ is one-hot encoded (i.e., only containing binary integer-valued labels, i.e., if particle $i$ is in internal state $a$, then $\left[\boldsymbol{p}_i^{(0)}\right]_b = \delta_{ab}$ for any internal state label $b$), this is equivalent to the pseudo-likelihood maximization (PLM) algorithm in the inverse Ising problem,[59–62] or the masked likelihood maximization (MLM) in training large language models (LLMs).[63–65] Therefore, for MC-UCG in which discrete internal states are directly sampled, we can also use the ISC framework to optimize its forcefield with the following PLM loss:

$$S_{\mathrm{PLM}}(\boldsymbol{\theta}) = \mathbb{E}_{\boldsymbol{r}^n \sim p(\boldsymbol{r}^n)} \left[ -\sum_i \log p_{i,s_i}(\boldsymbol{\theta} | M(\boldsymbol{r}^n), \{s_{-i}\}) \right] \tag{35}$$

with the pseudo-likelihood given by

$$p_{i,s_i}(\boldsymbol{\theta}|M(\boldsymbol{r}^n),\{s_{-i}\}) = \sigma_{\beta,s_i}\left(h_{i,s_i} + \sum_{j\in N(i)} u_{ij,s_is_j}(r_{ij})\right) \tag{36}$$

We note that the ISC loss is dependent on the number of ISC layers used during inference of all-atom data. During simulation, the inference is done once per MD step; but during postprocessing of simulation frames, which are usually stored with much larger time intervals, multilayer ISC is needed. Furthermore, as each layer of ISC produces its own output, we can also require that different layers of output and require that each intermediate layer also match the original data label in an $K$-layer ISC. We call this training strategy **Multilayer Internal State Consistency (MISC)**. The **MISC loss** in this case is

$$S_{\mathrm{MISC}}^{(K)}(\boldsymbol{\theta}) = \mathbb{E}_{\boldsymbol{r}^n\sim p(\boldsymbol{r}^n)}\left[-\sum_{k=1}^{K} w_k \sum_i \boldsymbol{p}_i^{(0)}(\boldsymbol{r}^n)\cdot\log\boldsymbol{p}_i^{(k)}\big(\boldsymbol{\theta}\big|M(\boldsymbol{r}^n)\big)\right] \tag{37}$$

with $w_k$ as a weight for linear combination for the cross entropy out of layer $k$. Although it is also a tunable hyperparameter, we use a naïve initialization: $w_k = \frac{1}{m}, \forall k$.

### D. Practical training strategy with both UCG-REM and Multilayer Internal State Consistency (MISC)

Because the UCG-REM loss and the ISC loss represent the two components in the KL divergence of the joint distribution, one should train the model with both UCG-REM loss and MISC loss to fully express the properties in the CG model. However, there are significant differences in the training paradigms of these two loss functions. The UCG-REM loss, just like REM itself, requires UCG sampling at each iteration and thus may suffer from sampling nuances, such as sampling noise, ergodicity breaking, etc. ISC/MISC training, as a regression task, only uses all-atom data samples and does not require UCG sampling at each iteration.

These differences inspire different training scheme in forcefield training (Fig 2), including the following:

1. **(UCG)** Training with only UCG-REM loss,
2. **(ISC/MISC)** Training with only ISC/MISC loss,
3. **(Co-training)** Co-training by combining UCG-REM loss and ISC loss, and
4. **(PT-FT)** Pre-training with ISC/MISC loss, then fine-tuning with UCG-REM loss.

From our numerical experiments, PT-FT outperforms other training schemes and is most insensitive to changes in training hyperparameter selection. It comprises of 3 training stages shown in Fig. 2(a): (1) REM without UCG, for initial guess; (2) ISC/MISC training, as the most important training step in recovering coupling in the UCG forcefield; (3) UCG-REM fine-tuning, to calibrate key observables.

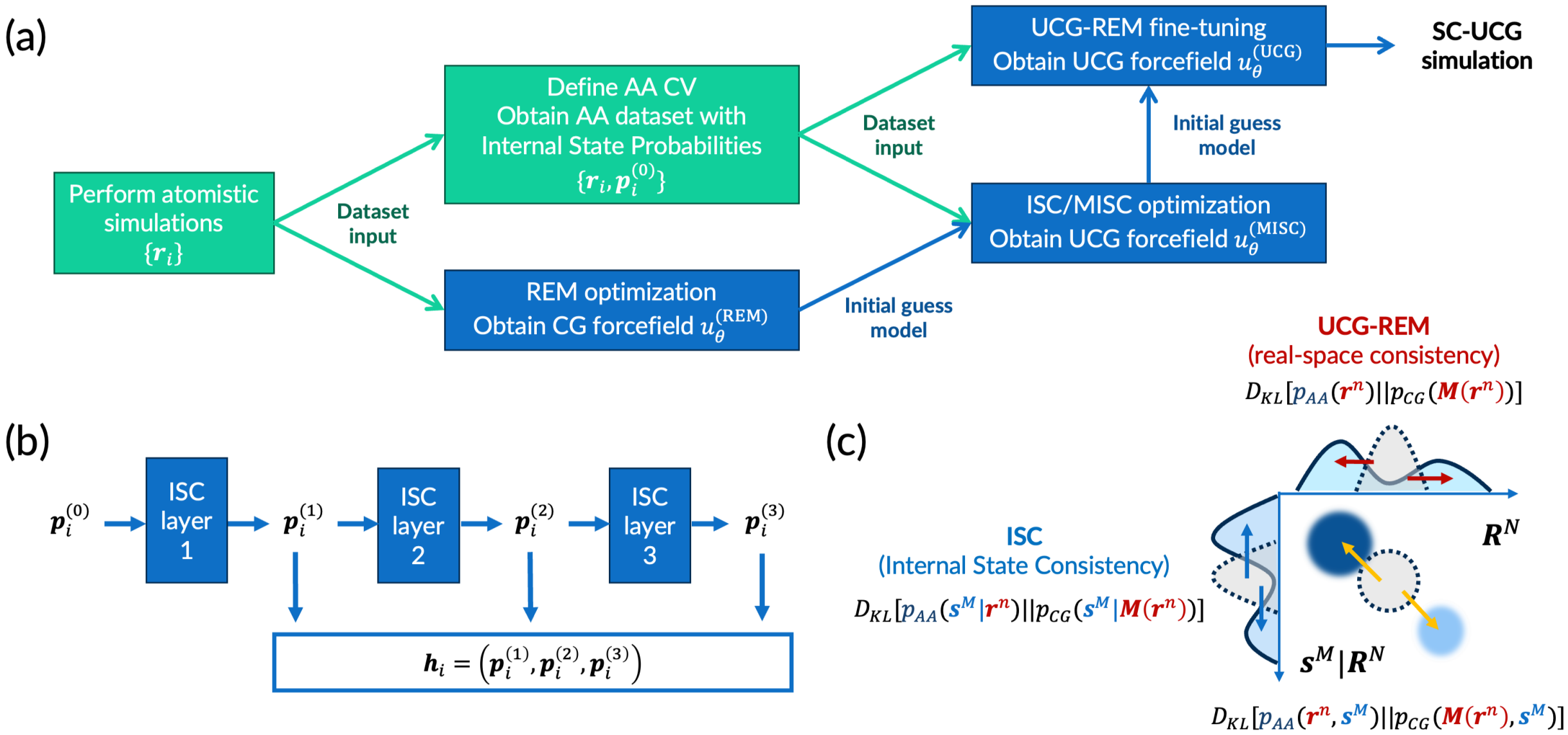

**FIG. 2. Overview of SC-UCG training methodology.** (a) Flowchart for the SC-UCG workflow for constructing an SC-UCG forcefield. Green boxes and arrows denote data construction steps and input direction; blue boxes and arrow denote model optimization. (b) Multilayer Internal State Consistency (MISC) training by concatenating outputs from different layers. (c) UCG-REM fine-tuning and ISC can be understood in terms of optimizing marginal and conditional distributions, respectively, of the joint distribution in the real space and the internal state space.

# III. Applications

## A. Dataset Preparation

To test the SC-UCG workflow on an exceptionally challenging system for a one bead CG model to capture, we chose a model tetramer system with chiral-racemic transitions in its fluid phase[41] (Fig. 3). This is a fluid system comprised of model molecules, each containing one dihedral angle subject to the following torsional energy:

$$U_{tors}(\{\boldsymbol{r}_i\}) = k_d \cos^2 \phi(\{\boldsymbol{r}_i\}) \tag{38}$$

which has two minima at $\phi = \pm\frac{\pi}{2}$, corresponding to the (D)- and (L)- enantiomers, respectively. The authors defined the chirality parameter for each molecule as following to determine their chiral states:

$$\zeta(\{\boldsymbol{r}_i\}) = -\frac{\boldsymbol{r}_{i,12} \cdot (\boldsymbol{r}_{i,23} \times \boldsymbol{r}_{i,34})}{|\boldsymbol{r}_{i,12}| \cdot |\boldsymbol{r}_{i,23}| \cdot |\boldsymbol{r}_{i,34}|} \in [-1, 1] \tag{39}$$

In our UCG ensemble (Fig. 3), we map the entire tetramer molecule in a single UCG site. We use $\{\boldsymbol{r}_i\}$ to denote the collection of 4 all-atom positions that correspond to CG site $i$. For each individual molecule, the sampled chirality parameter peaks sharply around -1 and 1, so we can directly define the dividing surface of L state (UCG state 0) and D state (UCG state 1) at $\zeta = 0$, and

$$p_{AA}(s_i = 1|\{\boldsymbol{r}_i\}) = \frac{1 + \zeta(\{\boldsymbol{r}_i\})}{2} \tag{40}$$

$$p_{AA}(s_i = 0|\{\boldsymbol{r}_i\}) = \frac{1 - \zeta(\{\boldsymbol{r}_i\})}{2} \tag{41}$$

In the all-atom system, the correlation of chirality is introduced by the intermolecular interaction between atoms $i$ and $j$ of different molecules. The effective Lennard-Jones interaction strength depends on the chirality of two molecules $\{\boldsymbol{r}_i\}$ and $\{\boldsymbol{r}_j\}$ respectively.

$$u_{ij}(r_{ij}) = 4\varepsilon_0\left[1 + \lambda\zeta(\{\boldsymbol{r}_i\})\zeta(\{\boldsymbol{r}_j\})\right]\left(r_{ij}^{-12} - r_{ij}^{-6}\right) \tag{42}$$

For this system, the reduced Lennard-Jones units are used throughout the paper. It has been numerically determined that, given the forcefield parameters, the critical temperature is $k_B T_c = 4.365\varepsilon_0$ (or, in reduced LJ units, $T_c = 4.365$). The coupling strength was chosen as $\lambda = 0.5$. Therefore, $T = 4.3$ is in the subcritical region, containing largely of chiral phases but exhibiting large spatial and temporal fluctuations in chirality. With a simulation box of $N = 1000$, we were able to sample infrequent transitions between L-dominant phases and D-dominant phases across the system. We extracted 500 frames (1 frame every 50 steps) from a 25,000 segment of reference trajectory at $T = 4.3$, with cubic simulation box length $L = 20.8\sigma$ and molecule number $N = 1000$ as the all-atom dataset.

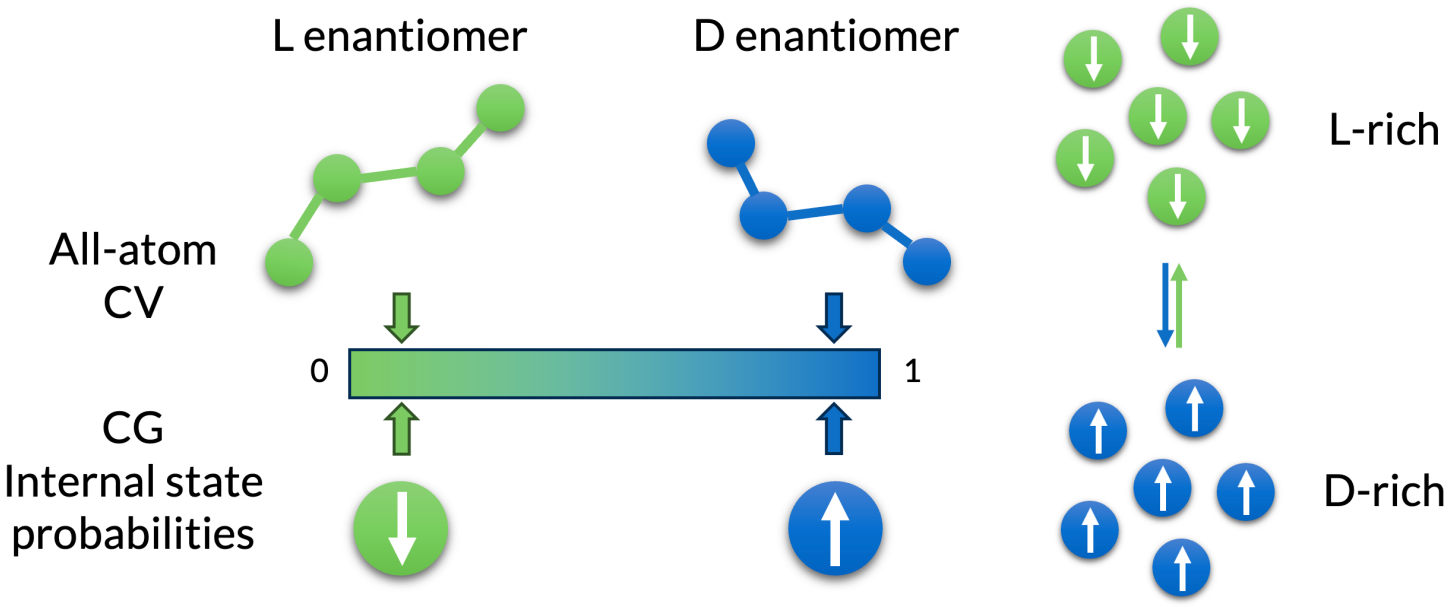

**FIG. 3.** Illustration of the chiral transition system as application.

### B. REM training and CG simulation parameters

For REM training, the neighbor list cutoff was set at $10\sigma$. The initial guess forcefield was obtained from direct Boltzmann inversion of the total pair distribution function. The repulsive core of the potential was modeled with a Gaussian peak with $x_0 = 0$ and maximum peak height $A = 100\varepsilon$. 3rd order B-splines were used with 100 coefficients. The Adam optimizer[66] was used with a learning rate of 0.01. The optimization converged within 100 iterations. The first 5 and last 5 knots of the forcefields were fixed with their initial values.

During the CG simulations of REM training, reduced Lennard-Jones units were applied, and CG simulation was conducted with timestep $\Delta t = 10^{-3}$. Neighbor list radius skin was set to $0.3\sigma$, smaller than the all-atom simulation, because the neighbor list is already large at $10\sigma$. Each REM iteration used $2 \times 10^5$ timesteps in total. All simulations were conducted in the constant $NVT$ ensemble with a Langevin thermostat having a damping timescale of 0.05.

### C. Multilayer ISC training

The UCG interaction matrix $\boldsymbol{u}_{ij}(r)$ was constructed as

$$\boldsymbol{u}_{ij}(r) = \begin{pmatrix} U_{\text{syn}}(r) & U_{\text{anti}}(r) \\ U_{\text{anti}}(r) & U_{\text{syn}}(r) \end{pmatrix} \tag{43}$$

With the D-D and L-L interactions identical to preserve the symmetry implied in the all-atom Hamiltonian. Both $U_{\mathrm{syn}}(r)$ and $U_{\mathrm{anti}}(r)$ functions were modeled also by 3rd order B-spline functions with 100 coefficients to optimize.

For Multilayer ISC (MISC) training, the forcefield obtained from REM was used as input unless otherwise noted for both. Cutoff radius for these interactions was varied between $6\sigma$, $8\sigma$ and $10\sigma$. The number of ISC layers were varied among 2, 3, 4. $K$-layer ISC training (taking outputs only from the $K$-th layer for the loss function) and were also conducted in a similar manner, with $K$ varied among 1, 2, 3, to probe their effects. Separate models were obtained for the mean-field and Bethe methods, and for different cutoff radii. Training was constructed with TorchREM, a PyTorch[67,68]-based package developed in this work that enables auto-differentiation during ISC/MISC training. The number of maximum epochs was 500, at which we already observed stable convergence in the ISC/MISC loss and accuracy. Each training epoch consumed less than 2 minutes on an NVIDIA GeForce 1080Ti graphics card.

### D. UCG-REM fine-tuning and SC-UCG simulation parameters

For each forcefield obtained from MISC training, a short UCG-REM fine-tuning was conducted from Iteration 400 at which the optimization had already converged. The Adam optimizer was applied with a learning rate of 0.001. The cutoff radii and B-spline coefficient numbers were maintained the same with the upstream MISC training.

The SC-UCG simulations were implemented and carried out with a separate UCG package in the LAMMPS MD package[69]. Each SC-UCG iteration used $2 \times 10^5$ steps. Internal state probability updates were conducted once per MD step. All other MD simulation parameters were the same as in the UCG-REM step.

## IV. Results and Discussion

Phase transitions are the most prevalent correlation-induced phenomena in condensed phase chemical physics. In our test UCG system, the chiral-racemic phase transition happens in the internal state space. To demonstrate the applicability of SC-UCG with correlated internal states, we were interested to test SC-UCG in the following two scenarios: (a) Near criticality, where large scale fluctuations should appear across the simulation box due to limited size effects. Especially at subcritical temperatures, there should be collective transitions between two chiral phases. Successful implementations of SC-UCG should capture such collective transitions. (b) Changing temperatures near criticality drastically alters the sampled distribution of individual chirality $\{\zeta_{i,t}\}$ and system-averaged chirality $\{\langle\zeta\rangle_t\}$. Since SC-UCG contains temperature in the message construction, we anticipate that, by training at a subcritical dataset only, the resulting CG model should then be capable of reproducing the phase transition in internal state space (i.e, to be "generative").

Furthermore, a successful UCG training strategy should be easy to control, monitor, and insensitive to hyperparameter changes. We also looked at how the fluctuation and phase transition behavior evolve during the UCG-REM fine-tuning iterations, and whether different hyperparameters in training can reproduce the same phenomena.

### A. SC-UCG captures correlation-induced structural statistics

We first looked at the obtained UCG forcefield functions, ISC loss and the structural statistics, i.e., the radial distribution function of the system. **Figs. 4a** and **4b** shows the obtained forcefields from each stage of optimization, with both MF and Bethe models. Comparing across different stages of training, the main optimization is undoubtedly done in the MISC stage of training. The UCG-REM fine-tuning with Adam optimizer and learning rate of 0.001 only slightly optimizes the forcefield, resulting in only very minor shifts to the sampled forcefield. Comparing the MISC forcefields and REM forcefields, we see that MISC training is able to capture the near-range cooperative interaction between ISC states; however, we also notice that $U_{\mathrm{syn}}$ from roughly $4.5\sigma$ to $6.5\sigma$ is higher than $U_{\mathrm{anti}}$, implying a slight mid-range anti-cooperativity. This arises due to the existence of the minority isomer outside of the cooperative core of the majority isomer at subcriticality, and the cross-over from the cooperative range to the anti-cooperative range corroborates with the correlation length sampled from the all-atom trajectory at $4.8\sigma$. When we change the MISC layer from $K = 3$ to 2 or 4 and the cutoff from $8\sigma$ to $6\sigma$ or $10\sigma$, these trends are very similar, except that the decrease in cutoff will lead to a slightly more attractive interaction for $U_{\mathrm{syn}}$. This can be interpreted as a way to mitigate the loss of internal state correlation without the contribution of farther neighbors.

Comparing the Bethe and MF results, we see that the near-range cooperativity from the forcefields appears more pronounced in the MF model. This is because MF models assume decorrelation of individual ISPs by default; correspondingly, given the same set of correlation-inducing UCG couplings, the $\boldsymbol{p}_{ij}$ matrices from the Bethe model will exhibit larger pair correlation than the MF model. Therefore, when solving for the inverse problem of trying to fit interactions to data, the coupling deduced from the MF model will tend to be higher than that in the Bethe model. However, both belonging to the mean-field family, the forcefields deduced from MF and Bethe models are qualitatively similar.

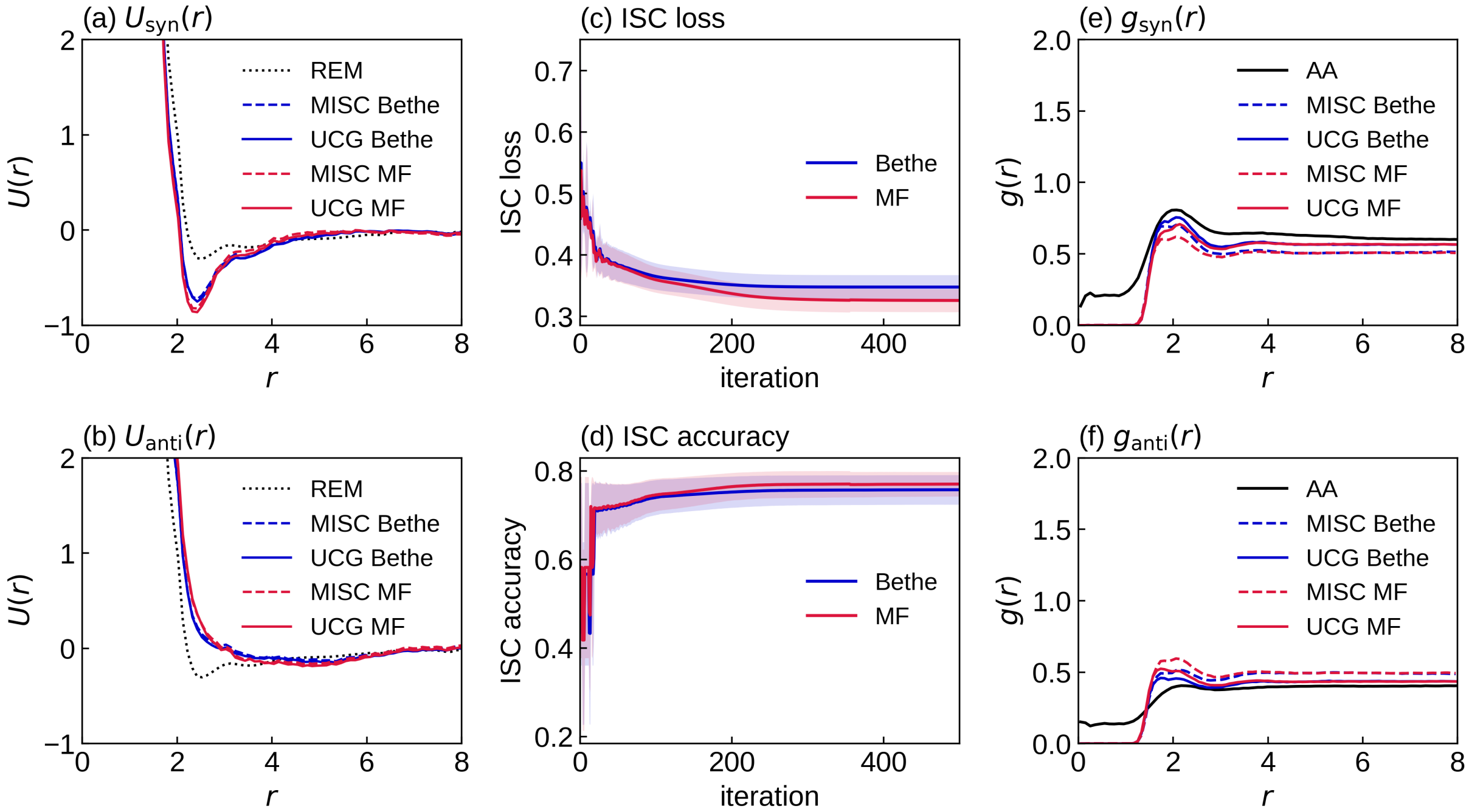

**FIG. 4. Forcefields, internal state classification, and state-wise distribution during SC-UCG training.** (a)-(b) Learned (a) homochiral (D-D and L-L) and (b) heterochiral (D-L) CG interaction $U_{\mathrm{syn}}$ after REM phase, MISC training phase (Iteration 400), and UCG-REM fine-tuning phase (Iteration 40), respectively. Inference layers were varied between mean-field (MF) and Bethe. (c) MISC loss, averaged by frame, during MISC training phase. The shaded regions represent (d) Internal state prediction accuracy during MISC training phase. (e) Internal-state-specific radial distribution function (ISS-RDF) for homochiral pairs (D-D

plus L-L). (f) ISS-RDF for heterochiral pairs (D-L). This figure is plotted for MISC layer $K = 3, r_c = 8\sigma$. For $K = 2, 4$ and $r_c = 6\sigma, 10\sigma$ results, see **Fig. S4.1-4.3** of Supporting Information.

During MISC training, we also monitored the evolution of the MISC loss and accuracy curves. The MISC loss converged beyond 400 iterations, and the training was very stable. To calculate accuracy, one should convert the ISP functions during sampling into discrete state labels. We directly attributed a CG site $i$ to be in state 1 if the output $p_{i,1} \geq 0.5$ and in state 0 if $p_{i,0} < 0.5$. The resulting accuracy curve converged cleanly after Iteration 250 without oscillatory behavior, stabilizing at around 0.75.

We also looked at the internal-state-specific radial distribution functions (ISS-RDF) during SC-UCG simulations, for syn- (D-D + L-L) and anti- (D-L) pairs. The ISS-RDF is calculated as follows:

$$g_{ab}(r) = \frac{1}{N_t} \sum_{t=0}^{N_t} \frac{2V}{N(N-1)} \sum_{(i,j)} \frac{\delta\left(r_{ij}(t) - r\right) p_{ij,ab}(r)}{4\pi r^2} \tag{44}$$

The syn- and anti- ISS-RDFs are defined by $g_{\text{syn}} = g_{00} + g_{11}$ and $g_{\text{anti}} = g_{01} + g_{10}$. For the all-atom trajectories, $p_{ij,s_i s_j}(r)$ is one-hot encoded. For the sampled CG trajectories, since we already have $p_{ij,s_i s_j}(r)$ sampled from MF or Bethe approximations for each pair during simulation, we directly use them as weights. The ISS-RDF $g_{ab}(r)$ does not converge to 1; rather, it converges to the total proportion of pairs (a) and (b) of all pairs. Therefore, it is a stricter metric for than the normalized RDF because large shifts in internal state correlations can drastically alter the relative heights of ISS-RDF. We can see from Figs. 4e and 4f that that the fine-tuned UCG-REM forcefield is able to provide an ISS-RDF that is better matched with the MISC forcefield. The repulsive part of the RDF is less aligned because of certain all-atom configurations that allow closer center-of-mass contacts, while the CG configuration is inherently hard-sphere. The Bethe approximation produces better-aligned ISS-RDFs to all-atom results due to its inclusion of pair correlation in the near distance; the MF potential, with stronger coupling learned and larger accuracy on single-site prediction, shows weaker pair correlation in the RDF. Therefore, without relying on defining any collective variable in the CG space, SC-UCG training and simulation-time inference can directly capture the structural statistics arising from internal state correlations in the all-atom dataset through the optimized forcefield parameters.

### B. SC-UCG captures correlation-induced collective fluctuations in internal states

To examine the ability of SC-UCG's to capture collective transitions between two chiral states in the subcritical region, we examined the distribution of enantiomer excess values sampled in the CG MD simulations during the UCG-REM fine-tuning, with different optimization hyperparameters. The enantiomer excess of one UCG particle $i$ at time $t$ is given directly by

$$ee_{i,t} = p_{i,1}(t) - p_{i,0}(t) = 2p_{i,1}(t) - 1 \tag{45}$$

and we denote the ensemble average of $\{ee_{i,t}\}$ at time $t$ as $\langle ee \rangle_t$. For each UCG-REM fine-tuning iteration, we examined the histogram for the distribution of $\langle ee \rangle_t$, $H(\langle ee \rangle_t)$, and the distribution of all sampled $\{ee_{i,t}\}$ across the system, $H(ee)$. We found that both MF and Bethe models can reproduce the collective fluctuations that are observed in the all-atom dataset, characterized by the bimodal distribution in $H(\langle ee \rangle_t)$, indicating system-wide chiral interconversion.

We also looked at how the MISC forcefields behave in different cutoff radius settings. The MISC forcefield in $r_c = 6$ produces a racemic result in which both $H(ee)$ and $H(\langle ee \rangle_t)$ are peaked at $ee = 0$ (**Figs. 5a and**

**5b**, Iteration 0). The MISC forcefield in $r_c = 8$ also had $H(ee)$ and $H(\langle ee \rangle_t)$ are peaked at $ee = 0$, but only 1 iteration of UCG-REM yields significant fluctuation into two energetically symmetric chiral states. The MISC forcefield in $r_c = 10$, however, have frequent fluctuations into the two chiral states (**Fig. 5g**). Although the distribution is still peaked at $ee = 0$, two chiral states have emerged as two smaller peaks in $H(\langle ee \rangle_t)$ at around $ee = \pm 0.35$. This reflects that the converged MISC forcefield yields forcefields sufficiently close to the critical point – but from above critical point. The temporal fluctuations in internal state space are enhanced if the neighbor list cutoff radius increases to near half of simulation box length.

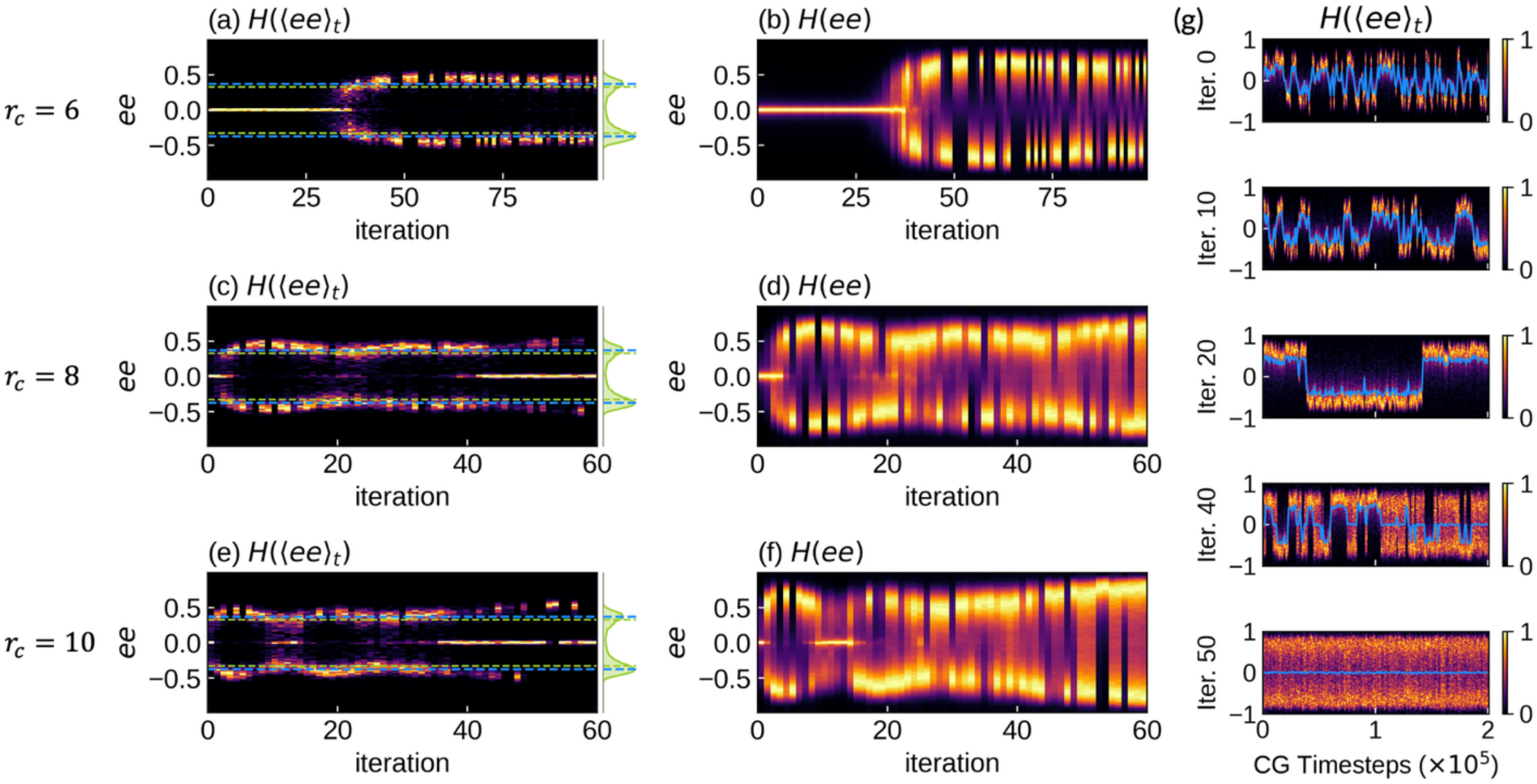

**FIG. 5. Emergence of collective chiral transitions during UCG-REM fine-tuning.** (a) Time histograms for the system average of enantiomer excess, $H(\langle ee \rangle_t)$, through UCG-REM iterations, starting from the MISC forcefield. The histogram densities were normalized with $p_{max} = 1$ (yellow) and $p_{min} = 0$ (black). The histograms were stacked horizontally to reflect per-iteration changes. The green dashed lines represent the time-averaged absolute value (symmetrized for better comparison) of enantiomer excess in the all-atom ensemble $\langle ee \rangle_{\mathrm{AA}}$. The blue dashed lines represent the most probable $\langle ee \rangle$ value in the symmetrized histogram from the all-atom data, $|ee|_{p,\mathrm{AA}}$. The green histograms on the right represent from the all-atom dataset at $T = 4.3$. (b) Histogram for the instantaneously sampled enantiomer excess for all particles at all sampled timesteps, $\{ee_{i,t}\}$, denoted as $H(ee)$, through UCG-REM iterations, starting from the MISC forcefield. In subplots (a)-(b), cutoff radius $r_c = 6\sigma$ for both MISC training and UCG-REM fine-tuning, and Bethe inference layers are used. (c)(d) Same as (a)-(b), except $r_c = 8\sigma$. (e)(f) Same as (a)-(b) except $r_c = 10\sigma$. (g) Time evolution of the normalized histogram of $ee$ values across system (heatmap), and the system average of enantiomer excess, $\langle ee \rangle_t$ (blue solid line), during the sampling phase for different UCG-REM iterations. Plotted for $r_c = 10\sigma$. For (a)-(f) mean-field results, see **SI Fig. S5.1**. For $r_c = 6\sigma, 8\sigma$ and mean-field results, see **SI Fig. S5.2**.

We then looked at how UCG-REM fine-tuning process trains the system into near-critical regions. For $r_c = 6$, no large temporal and spatial fluctuations of the sampled enantiomer excess values are observed until around Iteration 35, at which the system undergoes a bifurcation from the racemic state into two chiral states. During 100 steps of the fine-tuning process, most samples from UCG-REM exhibited only one uniform chiral state, although infrequent transitions are observed from Iterations 35 through 45, and some iterations after Iteration 75, as can be seen from the coexistence of two chiral peaks in Fig. 5b in these regions. For $r_c = 8$ and $r_c = 10$, it is much easier to obtain bimodal distribution in both $H(\langle ee \rangle_t)$ and $H(ee)$ through the first 50 iterations. Interestingly, we identify two "cross-over" regions around Iterations 20 and 40, respectively, for $r_c = 8$ (**Fig. 5c**). Transitions become much more frequent in this region, and the chiral peaks in $H(\langle ee \rangle_t)$ approach the all-atom system average $\langle |ee| \rangle_{AA}$ (green line in **Figs. 5a, 5c, 5e**) most closely. This "cross-over" behavior is the same in $r_c = 10$ but happens earlier, around Iterations 10 and 30 (**Fig. 5e**). A minor peak also appears near $ee = 0$ in these two "cross-over" regions, especially in the Bethe model, indicating the coexistence of both chiral states with a racemic state. This minor peak is less pronounced in the MF model (See **Supplementary Fig. S5.1**). In fact, in the subcritical region, the racemic state acts as the intermediate state between two racemic states. This is also echoed by the sampled all-atom distribution of $H(ee)$ with a minor peak centered around $ee = 0$. In those iterations between Iteration 0 and the first crossover, and between the two cross-over regions, the forcefield are slightly farther to the critical point as compared to the cross-over regions, with the transitions less frequent in sampling time. For example, in Fig. 5g, Iteration 20 exhibits less frequent transitions than Iterations 10 and 40. The second "cross-over" leads to a new region in training with $H(\langle ee \rangle_t)$ centered constantly again around 0 (with appearance of chiral peaks in some iterations), but $H(ee)$ remaining bimodal and peaking closer to $ee = \pm 1$. This indicates a region in which in every frame, the system has close to equal amounts of two chiral molecules, but their transitions are very fast across frames. This behavior is fundamentally different than the racemic behavior in Iteration 0, in which every particle exists in exactly $ee = \frac{1}{2}$ states and no spatial correlation can be explicitly recovered from the SC-UCG model. After the second cross-over event, the sampled distribution of $H(\langle ee \rangle_t)$ and $H(ee)$ from SC-UCG echoes that in a binary mixture, with each frame containing regions with different chirality, but no net chirality across the system.

In principle, by directly slowing down the rate of internal state updates and going to the slow-mixing regime, one would recover both temporal and spatial fluctuations in one sampling; but the slow-mixing regime can be directly sampled via MC-UCG without using the mean-field or Bethe approximations. Since our goal is to reproduce the collective transitions in the all-atom dataset by directly sampling the mixed-ensemble forcefield, without relying on tuning additional kinetic parameters, any forcefield in the two "cross-over" regions will already suffice for this purpose. Therefore, we conclude that SC-UCG is capable of capturing collective fluctuations in internal states due to correlations in the system.

### C. SC-UCG captures symmetry-breaking phase transition from single-temperature training

Since the SC-UCG forcefield is trained solely on the $T = 4.30$ dataset, we are interested in the behavior of the SC-UCG model when temperature is changed. With the forcefields obtained from the MISC training, we conducted simulations under different temperatures from $T = [4.2, 4.6]$, every 0.05 increment, to see if phase transition behavior can emerge. Figure 6 shows the sampled fluctuation of $\langle ee \rangle_t$ under different temperature settings. We see that at $T = 4.6$, the system is largely racemic with infrequent fluctuations into two chiral states. At $T = 4.5$ such fluctuations become increasingly frequent and at $T = 4.3$, symmetry breaking has developed, and we only see the racemic state appear as transient states in between the chiral transitions. At $T = 4.2$, the chiral transition slows down. This physical picture corroborates with the all-atom physical picture quantitatively, as most iterations inside two "cross-over" regions reproduce a critical transition within the $[4.2, 4.6]$ range. Compared to the all-atom results, such collective fluctuations happen at a faster rate, in line with the RLE premise of SC-UCG.

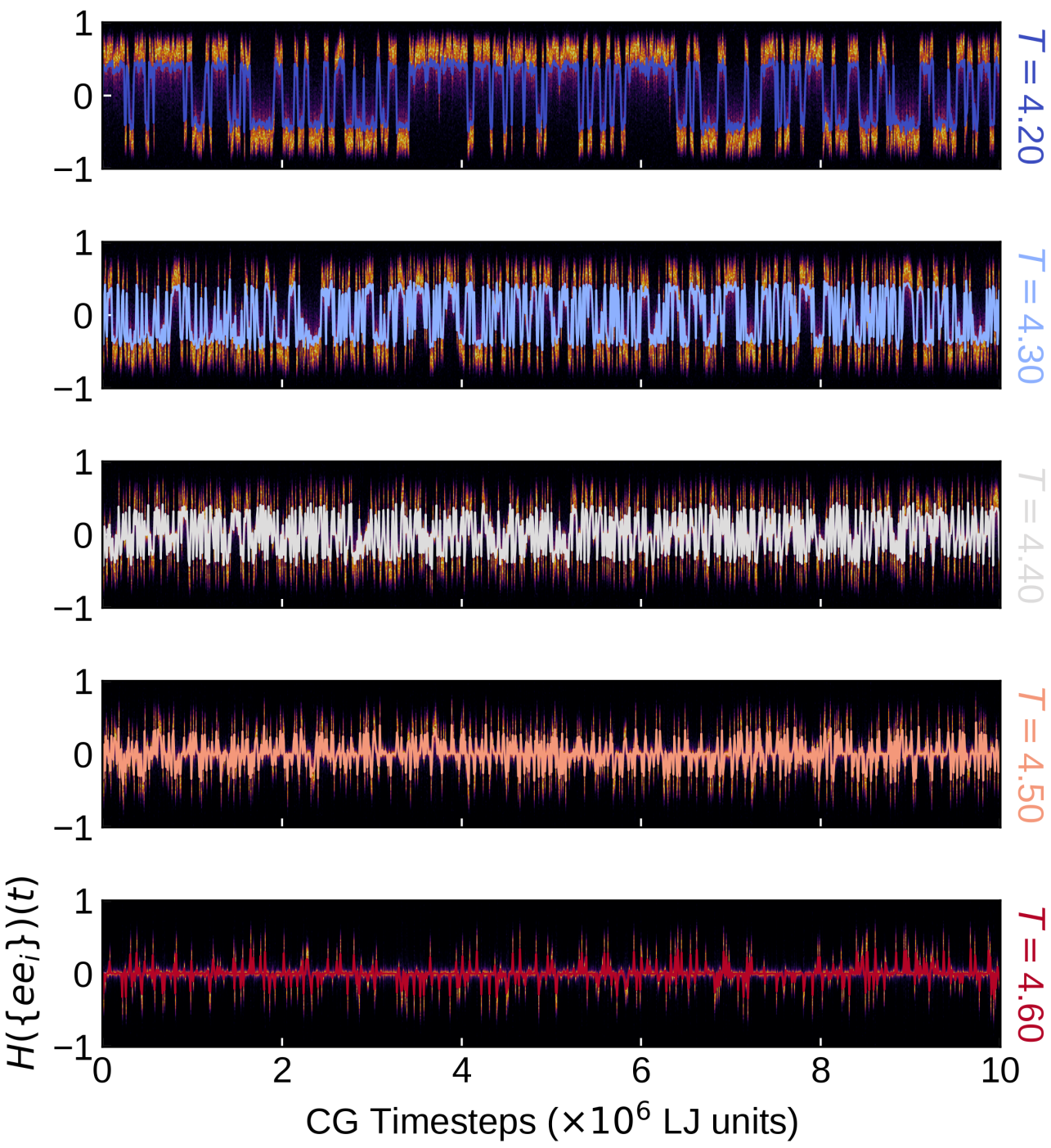

**FIG. 6.** Time evolution of the normalized histogram of $ee$ values across system (heatmap), and the system average of enantiomer excess, $\langle ee \rangle_t$ (blue solid line), for UCG-REM Iteration 20 of the Bethe model with 3-layer MISC and $r_c = 8$. For results from other iterations and other cutoff schemes, see SI Figs. S6.1-6.2.

Figure 7 shows the sampled histograms $H(\langle ee \rangle_t)$ and $H(ee)$ along different temperatures. By following the most probable system-average $ee$ value, $\langle |ee| \rangle_p$, and the most probable $ee$ values from all UCG particle, $|ee|_p$, from the two histograms respectively, we can determine the range of critical temperatures. We see that both Bethe and MF models can reproduce the critical temperature around $T = 4.4$. We note that near critical point, the behavior of the system is very sensitive to forcefield changes, and the shift in critical temperature can be drastic, even when the change in spline forcefield grids only happen in the order of $10^{-3}\ \varepsilon$. We also observe that the Bethe and MF methods can predict similar phase transition behavior. An interesting phenomenon is the difference of $|ee|_p(T)$ and observed at the end of the second "cross-over" during forcefield training. As the temperature increases, the $|ee|_p(T)$ transitions to $ee = 0$ at an earlier $T$. This phenomenon is observed to be more pronounced in the Bethe models: for example, in the UCG-REM Iteration 40 of 3-layer MISC with $r_c = 8\sigma$ (Fig. 7), $\langle |ee| \rangle_p(T)$ goes to zero at $T = 4.35$ while $|ee|_p(T)$ goes to zero at $T = 4.55$. In between these two points, beside the two chiral phases, the system takes on the rapid switching "chiral mixture" phase also observed in Fig. 5g, Iterations 40 and 50.

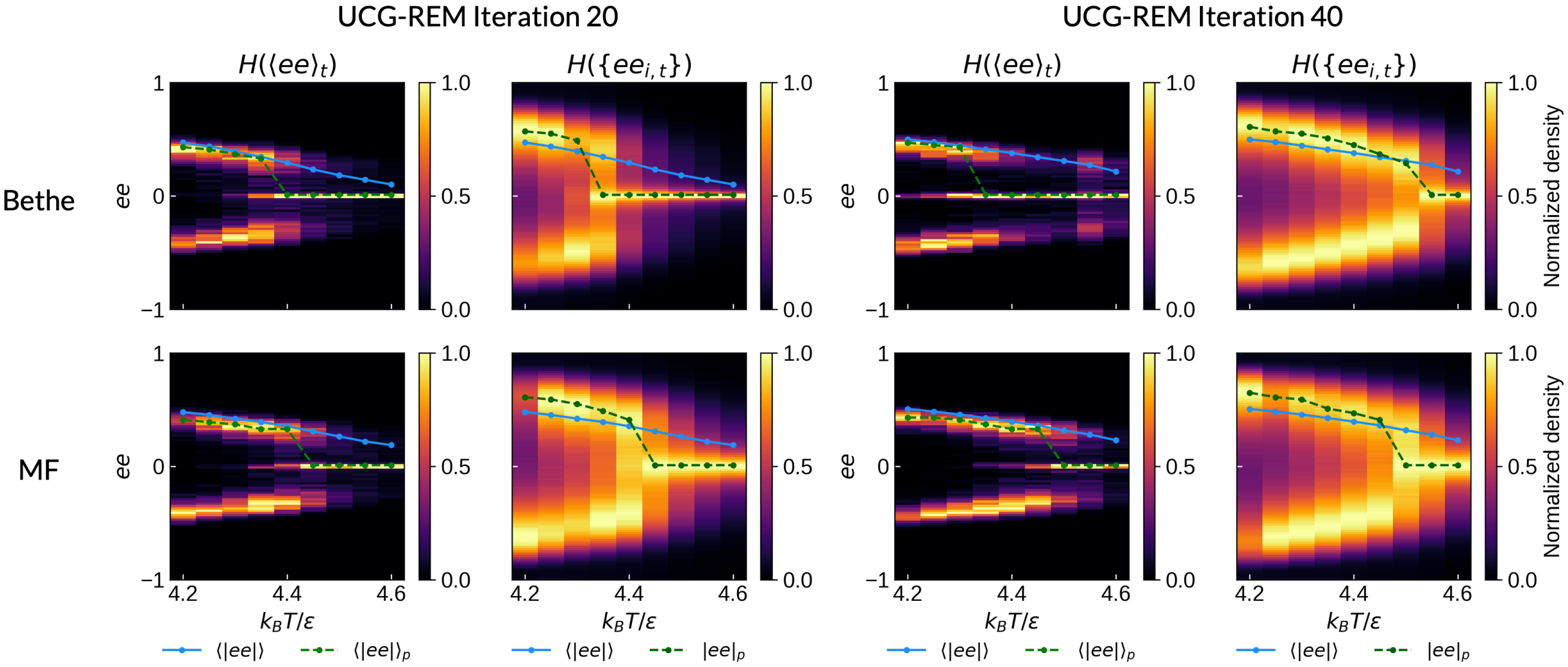

**FIG. 7. Phase transition in SC-UCG.** The heatmap in columns represent normalized histogram density (maximum density = 1) for each temperature. The forcefield is learned from the $T = 4.3$ dataset by PT-FT with 3-layer MISC at $r_c = 8\sigma$. For results from $r_c = 6\sigma$ and $r_c = 10\sigma$, see SI Figs S7.1. For other iterations from $r_c = 8\sigma$, see SI Figs. S7.2-7.3. Iteration 20 and 40 are arbitrarily chosen to represent the first and second "cross-over" regions, respectively. The blue solid line represents the ensemble average of $|ee|$; The green dashed line represents the most probable $\langle ee \rangle_t$ or $|ee|_{i,t}$ values from the symmetrized histograms.

Figure 8 compares the full-sized all-atom data from reference. Both the Bethe and MF data agree with the all-atom data qualitatively, and both models correctly capture the phase transition behavior similar to the all-atom scenario, but two major discrepancies are still present for future improvements. First, in the low-temperature regime, the extrapolation appears too mild, leading to smaller chirality than the all-atom dataset; second, in the high-temperature regime, the histograms heavily peaked around $ee = \frac{1}{2}$, denoting a strong mixed-state behavior and fast interconversion. These limitations mainly arise from single-temperature training. It is also important to note that the SC-UCG model is trained only from a down-sampled dataset, not from the monolithic full-sized all-atom dataset. Future extension of the MISC and UCG-REM training methods to a mixed multi-temperature ensemble is required to achieve full temperature transferability.

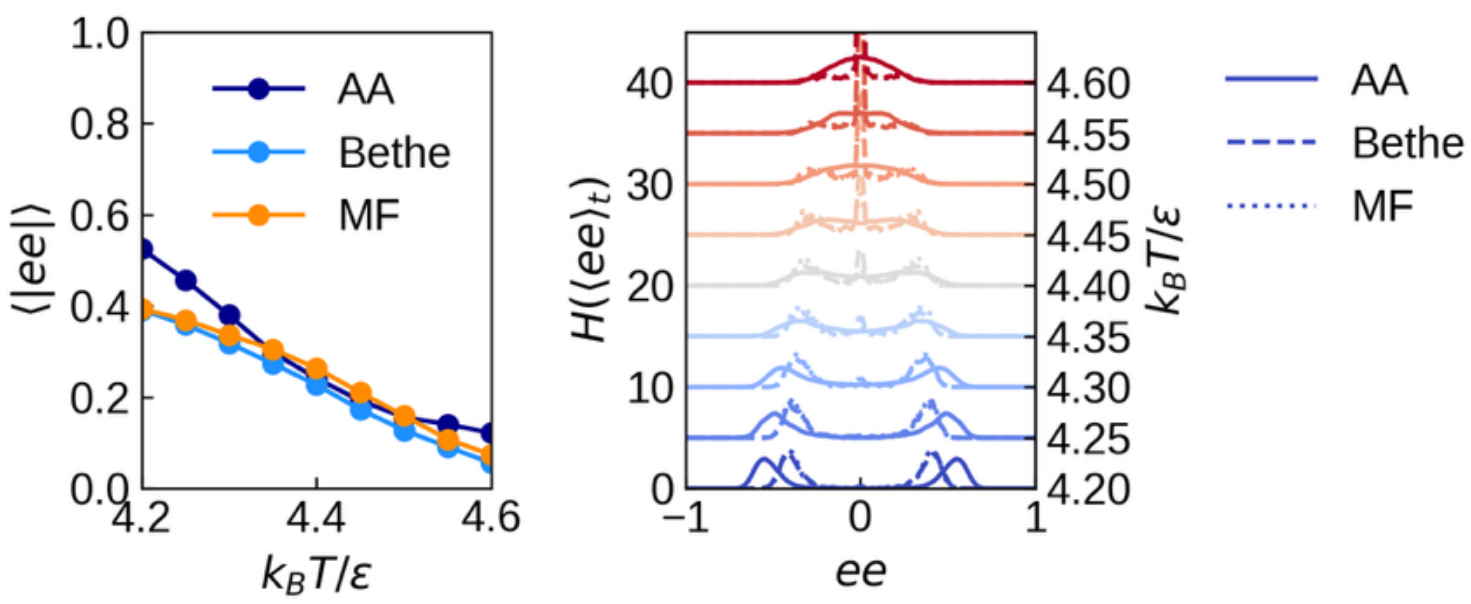

**FIG. 8. Comparison between the phase transition behavior in full-sized all-atom data and in SC-UCG.** The SC-UCG data is generated from 3-layer MISC training at $r_c = 8\sigma$. (a) Mean chirality in different temperatures from the AA dataset and the Bethe and MF variants of SC-UCG. (b) The enantiomer-excess histogram in different temperatures, compared between AA, Bethe, and MF.

## V. Conclusions and Outlook

In this paper, we have presented SC-UCG, a self-consistent inference method of internal states in bottom-up Ultra-Coarse-Graining (UCG). The approach introduces three new aspects of theory into UCG: (a) self-consistent updates to the internal state probability functions during simulation time, without relying on CG collective variables; (b) a generalized framework for including explicit pairwise correlation into the SC-UCG mixed Hamiltonian and the inference method, exemplified by the Bethe approximation; and (c) the Multilayer Internal State Consistency (MISC) training scheme, a systematic forcefield training scheme for SC-UCG, derived from the relative entropy minimization (REM) framework.

On the simulation side, SC-UCG relies on the premise of rapid local equilibrium (RLE), i.e., the UCG MD sampling is conducted with the effective Hamiltonian by approximately integrating out all internal states. This method also aligns with the MC-UCG idea that the internal states in are discrete-valued degrees of freedom, and its all-atom representation are collective variables of the mapped-out all-atom coordinates *within* the CG site to which they belong. By adopting a variational inference scheme, we showed that the minimization of the variational free energy naturally leads to a message-passing inference framework, with the UCG interaction matrix acting directly as the weight matrix for message construction. For the two-state UCG particle scenario, the Bethe approximation can also be employed to both the mixed SC-UCG Hamiltonian and the message-passing inference, with an additional edge update step.

Additionally, due to its inclusion of temperature in the internal state probability function, both the MF and Bethe models can describe temperature-dependent behavior of internal states for a given UCG forcefield. On the forcefield learning side, the ISC training scheme provides a robust workflow for constructing the SC-UCG forcefield, by employing a self-consistent cross-entropy loss over internal state probabilities across all UCG particles. Multilayer ISC concatenates output from different ISC layers and ensures self-consistency and robustness to training hyperparameters.

We applied SC-UCG to coarse-grain a model (but very non-trivial) chiral molecule system. Most importantly, we showed that SC-UCG reproduces the collective chiral transitions from the all-atom dataset, making SC-UCG suitable for simulating collective transitions. We showed that, even trained from a single-temperature dataset, SC-UCG is able to reproduce certain temperature-dependent properties regarding internal states, especially the chiral symmetry breaking phase transition. In terms of structural properties, the Bethe approximation captures the structural properties, such as the internal-state-specific radial distribution functions, better than the naïve MF approximation, although both do not achieve ideal matching of structural properties due to the inherent limitations of the mean-field family of approximations near the critical point, i.e., the lack of multi-body and longer-ranged internal state correlations.

The present method can be applied to a wide range of chemically and biochemically relevant systems as well. For example, we envision its further application in studying the difficult ripple phase behavior of lipid membrane bilayers.[26] For the chiral-racemic phase transition, we also expect SC-UCG to play a part in discovering more realistic biochemical systems that exhibit such symmetry breaking behavior, apart from synthesized liquid crystal-like molecules or model systems. With material constraints incorporated, SC-UCG can further be applied to a wide range of proton transport problems at the CG level.[70]

We also note several areas of future improvements and extensions of the current method. First, the SC-UCG workflow relies on pre-labeled internal states or probabilities. For more complex biomolecular systems, an important first step is to correctly discover the internal states from the unlabeled all-atom dataset. And consequently, it is important to design the CG mapping to preserve and effectively differentiate these different internal states. This facilitates the need for an automated internal state discovery and CG mapping scheme using machine-learning. Second, the SC-UCG approach can be interfaced with machine learning potentials (MLPs) or protein folding models.[63,71,72] By viewing atom/residue types as changeable "internal states" and allowing sequence mutation during simulation, we can use the UCG "internal state sampling" to sample sequence space jointly with real-space MD, to guide and evaluate molecule design. Third, the temperature transferability in SC-UCG from single-temperature training is still limited. It may be desirable to couple the SC-UCG method with multi-temperature training and dual-matching formalisms[73,74] to achieve full temperature transferability in UCG models, both in position space and in internal state space. It is also viable to utilize more advanced machine-learning techniques, such as symbolic regression, etc. to learn more complicated energy-related graph convolution operations during the internal state inference.

## Supplementary Material

See the supplementary material for comparisons of alternative training schemes (Figs. S1–S3); state-wise radial distribution functions and forcefields for additional MISC layer numbers $K = 2, 3, 4$ and cutoff radii $r_c = 6\sigma, 8\sigma, 10\sigma$ (Figs. S4.1–S4.3); the emergence of collective chiral transitions for the mean-field and Bethe treatments (Figs. S5.1–S5.2); the time evolution of internal-state fluctuation histograms (Figs. S6.1–S6.2); and additional phase-transition heatmaps across iterations and cutoff schemes (Figs. S7.1–S7.3).

## Acknowledgments

This material is based upon work supported by the National Science Foundation (NSF grant CHE-2102677). Simulations were performed using computing resources provided by the University of Chicago Research Computing Center (RCC).

## Author Declarations

### Conflict of Interest

The authors have no conflicts to disclose.

### Author Contributions

**Weizhi Xue**: Conceptualization (equal); Methodology; Software (lead); Numerical Experiments (lead); Writing – original draft.

**Xiao Shuai**: Software (preliminary) and Numerical Experiments (preliminary).

**Gregory A. Voth**: Conceptualization (equal); Supervision; Funding acquisition (lead); Writing – review & editing.

## Data Availability

The software packages used in all numerical experiments: TorchREM and UCG LAMMPS extension, are available on GitHub: https://github.com/KJAdams2000/LAMMPS-UCG-dev and https://github.com/KJAdams2000/TorchREM. Important source data and sample scripts are openly available on Zenodo (https://doi.org/10.5281/zenodo.21285638).

## References

[1] M. Karplus and J. A. McCammon, "Molecular dynamics simulations of biomolecules," Nat. Struct. Biol. **9**, 646–652 (2002).

[2] M. G. Saunders and G. A. Voth, "Coarse-Graining Methods for Computational Biology," Annu. Rev. Biophys. **42**, 73–93 (2013).

[3] A. J. Pak and G. A. Voth, "Advances in coarse-grained modeling of macromolecular complexes," Curr. Opin. Struct. Biol. **52**, 119–126 (2018).

[4] J. Jin, A. J. Pak, A. E. P. Durumeric, T. D. Loose, and G. A. Voth, "Bottom-up Coarse-Graining: Principles and Perspectives," J. Chem. Theory Comput. **18**, 5759–5791 (2022).

[5] W. G. Noid, "Perspective: Advances, Challenges, and Insight for Predictive Coarse-Grained Models," J. Phys. Chem. B **127**, 4174–4207 (2023).

[6] S. J. Marrink, H. J. Risselada, S. Yefimov, D. P. Tieleman, and A. H. De Vries, "The MARTINI Force Field: Coarse Grained Model for Biomolecular Simulations," J. Phys. Chem. B **111**, 7812–7824 (2007).

[7] P. C. T. Souza, R. Alessandri, J. Barnoud, S. Thallmair, I. Faustino, F. Grünewald, I. Patmanidis, H. Abdizadeh, B. M. H. Bruininks, T. A. Wassenaar, P. C. Kroon, J. Melcr, V. Nieto, V. Corradi, H. M. Khan, J. Domański, M. Javanainen, H. Martinez-Seara, N. Reuter, R. B. Best, I. Vattulainen, L. Monticelli, X. Periole, D. P. Tieleman, A. H. De Vries, and S. J. Marrink, "Martini 3: a general purpose force field for coarse-grained molecular dynamics," Nat. Methods **18**, 382–388 (2021).

[8] S. Izvekov and G. A. Voth, "A Multiscale Coarse-Graining Method for Biomolecular Systems," J. Phys. Chem. B **109**, 2469–2473 (2005).

[9] W. G. Noid, J.-W. Chu, G. S. Ayton, V. Krishna, S. Izvekov, G. A. Voth, A. Das, and H. C. Andersen, "The multiscale coarse-graining method. I. A rigorous bridge between atomistic and coarse-grained models," J. Chem. Phys. **128**, 244114 (2008).

[10] W. G. Noid, P. Liu, Y. Wang, J.-W. Chu, G. S. Ayton, S. Izvekov, H. C. Andersen, and G. A. Voth, "The multiscale coarse-graining method. II. Numerical implementation for coarse-grained molecular models," J. Chem. Phys. **128**, 244115 (2008).

[11] A. K. Soper, "Empirical potential Monte Carlo simulation of fluid structure," Chem. Phys. **202**, 295–306 (1996).

[12] T. C. Moore, C. R. Iacovella, and C. McCabe, "Derivation of coarse-grained potentials via multistate iterative Boltzmann inversion," J. Chem. Phys. **140**, 224104 (2014).

[13] M. S. Shell, "The relative entropy is fundamental to multiscale and inverse thermodynamic problems," J. Chem. Phys. **129**, 144108 (2008).

[14] A. Chaimovich and M. S. Shell, "Relative entropy as a universal metric for multiscale errors," Phys. Rev. E **81**, 060104 (2010).

[15] A. Chaimovich and M. S. Shell, "Coarse-graining errors and numerical optimization using a relative entropy framework," J. Chem. Phys. **134**, 094112 (2011).

[16] E. Pretti and M. S. Shell, "A microcanonical approach to temperature-transferable coarse-grained models using the relative entropy," J. Chem. Phys. **155**, 094102 (2021).

[17] J. Jin, A. J. Pak, and G. A. Voth, "Understanding Missing Entropy in Coarse-Grained Systems: Addressing Issues of Representability and Transferability," J. Phys. Chem. Lett. **10**, 4549–4557 (2019).

[18] J. Jin, K. S. Schweizer, and G. A. Voth, "Understanding dynamics in coarse-grained models. I. Universal excess entropy scaling relationship," J. Chem. Phys. **158**, 034103 (2023).

[19] T. T. Foley, M. S. Shell, and W. G. Noid, "The impact of resolution upon entropy and information in coarse-grained models," J. Chem. Phys. **143**, 243104 (2015).

[20] K. M. Kidder and W. G. Noid, "Analysis of mapping atomic models to coarse-grained resolution," J. Chem. Phys. **161**, 134113 (2024).

[21] N. J. H. Dunn, T. T. Foley, and W. G. Noid, "Van der Waals Perspective on Coarse-Graining: Progress toward Solving Representability and Transferability Problems," Acc. Chem. Res. **49**, 2832–2840 (2016).

[22] J. W. Wagner, J. F. Dama, A. E. P. Durumeric, and G. A. Voth, "On the representability problem and the physical meaning of coarse-grained models," J. Chem. Phys. **145**, 044108 (2016).

[23] J. F. Dama, A. V. Sinitskiy, M. McCullagh, J. Weare, B. Roux, A. R. Dinner, and G. A. Voth, “The Theory of Ultra-Coarse-Graining. 1. General Principles,” J. Chem. Theory Comput. **9**, 2466–2480 (2013).
[24] A. Davtyan, J. F. Dama, A. V. Sinitskiy, and G. A. Voth, “The Theory of Ultra-Coarse-Graining. 2. Numerical Implementation,” J. Chem. Theory Comput. **10**, 5265–5275 (2014).
[25] J. F. Dama, J. Jin, and G. A. Voth, “The Theory of Ultra-Coarse-Graining. 3. Coarse-Grained Sites with Rapid Local Equilibrium of Internal States,” J. Chem. Theory Comput. **13**, 1010–1022 (2017).
[26] P. G. Sahrmann and G. A. Voth, “Entropy-based methods for formulating bottom-up ultra-coarse-grained models,” J. Chem. Phys. **162**, 044102 (2025).
[27] T. Mori, R. J. Hamers, J. A. Pedersen, and Q. Cui, “Integrated Hamiltonian Sampling: A Simple and Versatile Method for Free Energy Simulations and Conformational Sampling,” J. Phys. Chem. B **118**, 8210–8220 (2014).
[28] Y. Q. Gao, “An integrate-over-temperature approach for enhanced sampling,” J. Chem. Phys. **128**, 064105 (2008).
[29] L. Yang and Y. Q. Gao, “A selective integrated tempering method,” J. Chem. Phys. **131**, 214109 (2009).
[30] X. Lin, Y. Xia, J. Zhang, and Y. Q. Gao, “Integrated Boltzmann Sampling: A Few-State Approach for Efficient Multistate Free Energy Calculations,” J. Chem. Theory Comput., acs.jctc.5c01240 (2026).
[31] S. Mani, H. H. Katkar, and G. A. Voth, “Compressive and Tensile Deformations Alter ATP Hydrolysis and Phosphate Release Rates in Actin Filaments,” J. Chem. Theory Comput. **17**, 1900–1913 (2021).
[32] J. M. A. Grime, J. F. Dama, B. K. Ganser-Pornillos, C. L. Woodward, G. J. Jensen, M. Yeager, and G. A. Voth, “Coarse-grained simulation reveals key features of HIV-1 capsid self-assembly,” Nat. Commun. **7**, 11568 (2016).
[33] M. Gupta, A. J. Pak, and G. A. Voth, “Critical mechanistic features of HIV-1 viral capsid assembly,” Sci. Adv., eadd7434 (2023).
[34] P. A. Pearce and H. L. Scott, “Statistical mechanics of the ripple phase in lipid bilayers,” J. Chem. Phys. **77**, 951–958 (1982).
[35] W. S. McCullough, J. H. H. Perk, and H. L. Scott, “Analysis of a model for the ripple phase of lipid bilayers,” J. Chem. Phys. **93**, 6070–6080 (1990).
[36] O. Lenz and F. Schmid, “Structure of Symmetric and Asymmetric ‘Ripple’ Phases in Lipid Bilayers,” Phys. Rev. Lett. **98**, 058104 (2007).
[37] M. Davies, A. D. Reyes-Figueroa, A. A. Gurtovenko, D. Frankel, and M. Karttunen, “Elucidating lipid conformations in the ripple phase: Machine learning reveals four lipid populations,” Biophys. J. **122**, 442–450 (2023).
[38] C. Dressel, T. Reppe, M. Prehm, M. Brautzsch, and C. Tschierske, “Chiral self-sorting and amplification in isotropic liquids of achiral molecules,” Nat. Chem. **6**, 971–977 (2014).
[39] T. Reppe, S. Poppe, X. Cai, Y. Cao, F. Liu, and C. Tschierske, “Spontaneous mirror symmetry breaking in benzil-based soft crystalline, cubic liquid crystalline and isotropic liquid phases,” Chem. Sci. **11**, 5902–5908 (2020).
[40] Y. Wang, F. H. Stillinger, and P. G. Debenedetti, “Fluid–fluid phase transitions in a chiral molecular model,” J. Chem. Phys. **157**, 084501 (2022).
[41] P. M. Piaggi, R. Car, F. H. Stillinger, and P. G. Debenedetti, “Critical behavior in a chiral molecular model,” J. Chem. Phys. **159**, 114502 (2023).
[42] R. Menta, P. G. Debenedetti, R. Car, and P. M. Piaggi, “On the Possibility of Chiral Symmetry Breaking in Liquid Hydrogen Peroxide,” J. Phys. Chem. B **129**, 5335–5342 (2025).
[43] J. Jin and G. A. Voth, “Ultra-Coarse-Grained Models Allow for an Accurate and Transferable Treatment of Interfacial Systems,” J. Chem. Theory Comput. **14**, 2180–2197 (2018).
[44] K. T. Schütt, P.-J. Kindermans, H. E. Sauceda, S. Chmiela, A. Tkatchenko, and K.-R. Müller, “SchNet: A continuous-filter convolutional neural network for modeling quantum interactions,” arXiv:arXiv:1706.08566 (2017).
[45] T. D. Loose, P. G. Sahrmann, T. S. Qu, and G. A. Voth, “Coarse-Graining with Equivariant Neural Networks: A Path Toward Accurate and Data-Efficient Models,” J. Phys. Chem. B **127**, 10564–10572 (2023).

[46] I. Batatia, D. P. Kovács, G. N. C. Simm, C. Ortner, and G. Csányi, “MACE: Higher Order Equivariant Message Passing Neural Networks for Fast and Accurate Force Fields,” arXiv:arXiv:2206.07697 (2023).
[47] B. Cseke and T. Heskes, “Properties of Bethe Free Energies and Message Passing in Gaussian Models,” J. Artif. Intell. Res. **41**, 1–24 (2011).
[48] A. Weller, K. Tang, D. Sontag, and T. Jebara, “Understanding the Bethe approximation: when and how can it go wrong?,” in *Proc. Thirtieth Conf. Uncertain. Artif. Intell.* (AUAI Press, Arlington, Virginia, USA, 2014), pp. 868–877.
[49] L. Wu, P. Cui, J. Pei, and L. Zhao, editors , *Graph Neural Networks: Foundations, Frontiers, and Applications* (Springer Nature Singapore, Singapore, 2022).
[50] M. Ahsan, C. Pindi, S. Sinha, A. C. Patel, and G. Palermo, “Graph neural networks for molecular dynamics simulations,” Curr. Opin. Struct. Biol. **97**, 103238 (2026).
[51] H. Leisenberger and F. Pernkopf, “Adaptive Variational Inference in Probabilistic Graphical Models: Beyond Bethe, Tree-Reweighted, and Convex Free Energies,” arXiv:arXiv:2502.03341 (2025).
[52] A. Pelizzola, “Cluster Variation Method in Statistical Physics and Probabilistic Graphical Models,” J. Phys. Math. Gen. **38**, R309–R339 (2005).
[53] S. Cocco and R. Monasson, “Adaptive Cluster Expansion for Inferring Boltzmann Machines with Noisy Data,” Phys. Rev. Lett. **106**, 090601 (2011).
[54] S. Cocco and R. Monasson, “Adaptive Cluster Expansion for the Inverse Ising Problem: Convergence, Algorithm and Tests,” J. Stat. Phys. **147**, 252–314 (2012).
[55] J. P. Barton, E. De Leonardis, A. Coucke, and S. Cocco, “ACE: adaptive cluster expansion for maximum entropy graphical model inference,” Bioinformatics **32**, 3089–3097 (2016).
[56] W. Wiegerinck and T. Heskes, “Fractional belief propagation,” in *Proc. 16th Int. Conf. Neural Inf. Process. Syst.* (MIT Press, Cambridge, MA, USA, 2002), pp. 438–445.
[57] P. Veličković, G. Cucurull, A. Casanova, A. Romero, P. Liò, and Y. Bengio, “Graph Attention Networks,” arXiv:arXiv:1710.10903 (2018).
[58] C. Hua, S. Luan, Q. Zhang, and J. Fu, “Graph Neural Networks Intersect Probabilistic Graphical Models: A Survey,” arXiv:arXiv:2206.06089 (2023).
[59] A. Decelle and F. Ricci-Tersenghi, “Pseudolikelihood Decimation Algorithm Improving the Inference of the Interaction Network in a General Class of Ising Models,” Phys. Rev. Lett. **112**, 070603 (2014).
[60] M. Ekeberg, C. Lövkvist, Y. Lan, M. Weigt, and E. Aurell, “Improved contact prediction in proteins: Using pseudolikelihoods to infer Potts models,” Phys. Rev. E **87**, 012707 (2013).
[61] E. Aurell and M. Ekeberg, “Inverse Ising Inference Using All the Data,” Phys. Rev. Lett. **108**, 090201 (2012).
[62] S. Cocco, C. Feinauer, M. Figliuzzi, R. Monasson, and M. Weigt, “Inverse statistical physics of protein sequences: a key issues review,” Rep. Prog. Phys. **81**, 032601 (2018).
[63] T. Hayes, R. Rao, H. Akin, N. J. Sofroniew, D. Oktay, Z. Lin, R. Verkuil, V. Q. Tran, J. Deaton, M. Wiggert, R. Badkundri, I. Shafkat, J. Gong, A. Derry, R. S. Molina, N. Thomas, Y. A. Khan, C. Mishra, C. Kim, L. J. Bartie, M. Nemeth, P. D. Hsu, T. Sercu, S. Candido, and A. Rives, “Simulating 500 million years of evolution with a language model,” Science **387**, 850–858 (2025).
[64] P. Moreno-Muñoz, P. G. Recasens, and S. Hauberg, “On Masked Pre-training and the Marginal Likelihood,” arXiv:arXiv:2306.00520 (2023).
[65] R. Rende, F. Gerace, A. Laio, and S. Goldt, “Mapping of attention mechanisms to a generalized Potts model,” Phys. Rev. Res. **6**, 023057 (2024).
[66] D. P. Kingma and J. Ba, “Adam: A Method for Stochastic Optimization,” arXiv:arXiv:1412.6980 (2017).
[67] A. Paszke, S. Gross, F. Massa, A. Lerer, J. Bradbury, G. Chanan, T. Killeen, Z. Lin, N. Gimelshein, L. Antiga, A. Desmaison, A. Köpf, E. Yang, Z. DeVito, M. Raison, A. Tejani, S. Chilamkurthy, B. Steiner, L. Fang, J. Bai, and S. Chintala, “PyTorch: An Imperative Style, High-Performance Deep Learning Library,” arXiv:arXiv:1912.01703 (2019).
[68] M. Fey and J. E. Lenssen, “Fast Graph Representation Learning with PyTorch Geometric,” arXiv:arXiv:1903.02428 (2019).

[69] A. P. Thompson, H. M. Aktulga, R. Berger, D. S. Bolintineanu, W. M. Brown, P. S. Crozier, P. J. In ’T Veld, A. Kohlmeyer, S. G. Moore, T. D. Nguyen, R. Shan, M. J. Stevens, J. Tranchida, C. Trott, and S. J. Plimpton, “LAMMPS - a flexible simulation tool for particle-based materials modeling at the atomic, meso, and continuum scales,” Comput. Phys. Commun. **271**, 108171 (2022).
[70] J. Jin, Z. Li, and G. A. Voth, “Systematic bottom-up coarse-graining of hydrated excess proton transport across scales,” Nat. Comput. Sci. (2026).
[71] J. B. Ingraham, M. Baranov, Z. Costello, K. W. Barber, W. Wang, A. Ismail, V. Frappier, D. M. Lord, C. Ng-Thow-Hing, E. R. Van Vlack, S. Tie, V. Xue, S. C. Cowles, A. Leung, J. V. Rodrigues, C. L. Morales-Perez, A. M. Ayoub, R. Green, K. Puentes, F. Oplinger, N. V. Panwar, F. Obermeyer, A. R. Root, A. L. Beam, F. J. Poelwijk, and G. Grigoryan, “Illuminating protein space with a programmable generative model,” Nature **623**, 1070–1078 (2023).
[72] J. Airas and B. Zhang, “Knowledge Distillation of a Protein Language Model Yields a Foundational Implicit Solvent Model,” arXiv:arXiv:2601.05388 (2026).
[73] K. M. Lebold and W. G. Noid, “Dual approach for effective potentials that accurately model structure and energetics,” J. Chem. Phys. **150**, 234107 (2019).
[74] E. Pretti and M. S. Shell, “A microcanonical approach to temperature-transferable coarse-grained models using the relative entropy,” J. Chem. Phys. **155**, 094102 (2021).